\documentclass[fleqn,usenatbib]{mnras}
\usepackage{newtxtext,newtxmath}
\usepackage[T1]{fontenc}

\DeclareRobustCommand{\VAN}[3]{#2}
\let\VANthebibliography\thebibliography
\def\thebibliography{\DeclareRobustCommand{\VAN}[3]{##3}\VANthebibliography}

\usepackage{graphicx}	%
\usepackage{amsmath}	%
\usepackage{tablefootnote} 

\usepackage{multicol}

\usepackage{xspace} %
\renewcommand{\AA}{\normalfont\r{A}\xspace} %

\usepackage{color,soul}
\definecolor{lightblue}{rgb}{.70,.95,1}
\sethlcolor{lightblue}

\newcommand{\teff}{\ensuremath{T_{\mathrm{eff}}}\xspace}

\newcommand{\kms}{\ensuremath{\rm{km}\,s^{-1}}\xspace}
\newcommand{\logg}{\ensuremath{\log g}\xspace}
\newcommand{\feh}{\rm{[Fe/H]}\xspace}
\newcommand{\fehphot}{\rm{[Fe/H]\ensuremath{_\mathrm{phot}}}\xspace}

\newcommand{\alphafe}{\rm{[\ensuremath{\alpha}/Fe]}\xspace}
\newcommand{\Gaia}{\textit{Gaia}\xspace}
\newcommand{\CaHK}{\emph{CaHK}\xspace}
\newcommand{\Pristine}{\emph{Pristine}\xspace}
\newcommand{\FERRE}{{\tt FERRE}\xspace}

\defcitealias{martin24}{M24}
\defcitealias{andrae23}{A23}

\title[PIGS photometry DR]{The Pristine Inner Galaxy Survey (PIGS) XII: Photometric data release and follow-up strategy for 4MOST}

\author[Ardern-Arentsen et al.]{Anke Ardern-Arentsen,$^{1,2}$\thanks{Email: \url{a.ardern-arentsen@surrey.ac.uk}}
Nicolas F. Martin,$^{3,4}$
Else Starkenburg,$^{5}$
Zhen Yuan,$^{6,7}$
\newauthor
Aleksandra Isajeva,$^{8}$
Cristina Chiappini,$^{9}$ 
Ivan Minchev,$^{9}$ 
Sébastien Fabbro,$^{10}$ 
Vanessa Hill,$^{11}$ 
\newauthor
Pascale Jablonka,$^{12}$  
Georges Kordopatis,$^{11}$ 
Julio F. Navarro,$^{13,14}$ 
Rubén Sánchez-Janssen,$^{15,16}$ 
\newauthor
Akshara Viswanathan$^{13}$ 
and Sara Vitali$^{16,17}$  \\
$^{1}$ School of Mathematics and Physics, University of Surrey, Guildford GU2 7XH, UK\\
$^{2}$ Institute of Astronomy, University of Cambridge, Madingley Road, Cambridge CB3 0HA, UK\\
$^{3}$ Universit\'e de Strasbourg, CNRS, Observatoire astronomique de Strasbourg, UMR 7550, F-67000 Strasbourg, France\\
$^{4}$ Max-Planck-Institut f\"ur Astronomie, K\"onigstuhl 17, D-69117 Heidelberg, Germany\\
$^{5}$ Kapteyn Astronomical Institute, University of Groningen, Postbus 800, 9700 AV, Groningen, the Netherlands\\
$^{6}$ School of Astronomy and Space Science, Nanjing University, Nanjing, Jiangsu 210093, China \\
$^{7}$ Key Laboratory of Modern Astronomy and Astrophysics, Nanjing University, Ministry of Education, Nanjing 210093, China\\
$^{8}$ Cavendish Laboratory, University of Cambridge, JJ Thomson Avenue, Cambridge CB3 0HE, UK \\
$^{9}$ Leibniz-Institut f\"ur Astrophysik Potsdam (AIP), An der Sternwarte 16, D-14482 Potsdam, Germany \\
$^{10}$ NRC Herzberg Astronomy and Astrophysics, 5071 West Saanich Road, Victoria, BC V9E 2E7, Canada \\
$^{11}$ Observatoire de la Côte d'Azur, Université Côte d'Azur, CNRS, Laboratoire Lagrange 06304 Nice, France \\
$^{12}$ Laboratoire d'astrophysique, École Polytechnique Fédérale de Lausanne (EPFL), CH-1290 Versoix, Switzerland \\
$^{13}$ Department of Physics and Astronomy, University of Victoria, PO Box 3055, STN CSC, Victoria, BC V8W 3P6, Canada \\
$^{14}$ Crimson Distinguished Professor, KU-KIST Green School, Korea University \\
$^{15}$ Isaac Newton Group of Telescopes, Apartado 321, E-38700 Santa Cruz de la Palma, Tenerife, Spain \\
$^{16}$ Instituto de Astrofísica de Canarias, C. Vía Láctea, s/n, 38205 La Laguna, Santa Cruz de Tenerife, Spain \\
$^{17}$ Universidad de La Laguna, Dpto. Astrofísica, Av. Astrofísico Francisco Sánchez, 38206 La Laguna, Santa Cruz de Tenerife, Spain
}

\date{Accepted XXX. Received YYY; in original form ZZZ}

\pubyear{2026}

\begin{document}
\label{firstpage}
\pagerange{\pageref{firstpage}--\pageref{lastpage}}
\maketitle

\begin{abstract}
Metal-poor stars in the inner Galaxy are ancient probes of the early Milky Way. 
We present the photometric data release of the Pristine Inner Galaxy Survey (PIGS), designed to identify metal-poor stars in this region.
PIGS is based on CFHT/MegaCam \CaHK imaging over $340\,\mathrm{deg}^2$ toward the Galactic bulge and the Sagittarius dwarf galaxy, and has been reprocessed and calibrated to the scale of the main \Pristine survey.
We release the \CaHK photometry for 55 million stars in both PIGS and \Gaia DR3, with typical \CaHK uncertainties $<0.1$~mag for $G<19.5$, and photometric metallicities for $\sim$36 million stars, of which $\sim$11 million pass basic quality cuts.
A comparison with spectroscopic metallicities shows that an empirical extinction rescaling is required to minimise photometric metallicity--extinction trends. 
After this correction, there is a good correlation between spectroscopic and photometric metallicities, with a systematic offset of $\sim 0.35\,\mathrm{dex}$ and a dispersion increasing from $\sim0.2$~dex for E(B$-$V) $<0.35$ to $0.3-0.45$~dex for E(B$-$V) up to $0.6-0.8$. 
Over 1.2 million stars with $G<17$ pass our stricter quality criteria, with $\sim20 \%, 1.4 \%$ and $ 0.04 \%$ having [Fe/H]$_\mathrm{phot}<-1$, $-2$, and $-3$, respectively (after the 0.35~dex correction).
Using high-resolution spectra of extremely metal-poor stars, we assess the potential effect of interstellar Ca H\&K on PIGS photometric metallicities and find typical metallicity biases of $\sim 0.05\,\mathrm{dex}$, reaching $\sim 0.2\,\mathrm{dex}$ in the strongest case. 
We present the  strategy for the 4MOST 4MIDABLE-LR/VMP\_PIGS sub-survey, which will result in $R\sim 6{,}500$ spectroscopy for $\sim 20{,}000$ inner-Galaxy stars with [Fe/H]$_\mathrm{phot} < -1.75$. This dataset will enable unique chemo-dynamical studies of the ancient inner Milky Way.
\end{abstract}


\section{Introduction}

In a hierarchical galaxy formation framework, the central few kiloparsecs of the Milky Way are expected to host our Galaxy's oldest stellar populations \citep[e.g.][]{white00,starkenburg17a, elbadry18}. The ancient stars in the inner Galaxy likely probe a variety of birth environments, from \emph{in-situ} to large early building blocks to smaller and larger accreted dwarf galaxy systems. The first long-lived stars that formed in these environments are the most metal-poor, which can be used to probe early Galaxy evolution and star formation \citep[see the review by][]{frebelnorris15}. Metal-poor bulge clusters ($-1.3 < \feh < -0.7$) are $\sim12.0-13.5$~Gyr old \citep{souza24, ortolani26}, so the very metal-poor ($\feh < -2.0$) in-situ phase must have been very short. 

The dominant structure in the inner Milky Way is the boxy/peanut-shaped bulge, which has relatively high metallicity ($-1.0 < \feh < +0.5$, e.g. \citealt{mcw94, ness13a}) and is thought to predominantly have formed from the Galactic disc through secular processes \citep[see the review by][]{barbuy18}. It is possible that there is also a more minor, spheroidal bulge component present at these metallicities \citep{zoccali24,nepal26}. 
The origin of the more metal-poor stars ($\feh < -1$) in the inner Galactic regions has been discussed much in recent years \citep[e.g.][]{ness13b, schiavon17, kunder20, horta21a, belokurovkravtsov22, razera22, lucey22, arentsen24, nepal26}, converging on a picture where these stars trace a halo-like population distinct from the disc-originated bulge, with some unique features compared to the rest of the halo that suggest potential formation in-situ, in globular clusters and/or in large Galactic building blocks. Very/extremely metal-poor stars ($\feh < -2.0/-3.0$, respectively) are very rare in the inner Galaxy, and larger samples and more work are needed to understand their origins. 

Dedicated searches are generally needed to build significant samples of the lowest metallicity stars, and in the inner Galaxy this is even more true due to their extreme scarcity. The search for low-metallicity stars is also hampered by the high amount of dust. As a community, we started braving the search for very metal-poor stars in this region over the past $\sim 10-15$ years, employing pre-selection techniques similar to those pioneered by the HK survey \citep{beers92} and the Hamburg-ESO survey \citep{christlieb08} in the (much-less-reddened) halo. Inner Galaxy searches have been based on narrow-band metallicity-sensitive photometry covering the Ca\,H\&K region (SkyMapper, \citealt{wolf18, howes14, lucey19}; \textit{Pristine}, \citealt{starkenburg17b, arentsen20_II}) or a metallicity-sensitive combination of infrared photometry (Best \& Brightest, B\&B, \citealt{schlaufmancasey14, caseyschlaufman15}). Very metal-poor candidates were selected for spectroscopic follow-up based on their location in colour-colour planes, with promising regions identified thanks to overlap with stars with available spectroscopy. None of these works produced photometric metallicities, due to challenges with dereddening and calibrating the photometry. Another strategy is to search for RR~Lyrae variables through their characteristic photometric variability, as inner-Galaxy RR~Lyrae predominantly trace an old, metal-poor population \citep{kunder22}; however, only a minority of them appear to be \emph{very} metal-poor \citep[e.g.][]{savino20}, and there are fewer RR Lyrae stars compared to red giant branch (RGB) stars.

In this paper, we present the first set of inner Galaxy photometric metallicities down to the very metal-poor regime, based on the \Pristine Inner Galaxy Survey \citep[PIGS;][]{arentsen20_II} narrow-band Canada-France-Hawaii Telescope (CFHT)/MegaCam \citep{boulade03} \CaHK photometry. A key enabler of this work is the availability of \Gaia DR3 XP spectro-photometry \citep{gaiadr3,deangeli23}, which allows us to homogeneously calibrate the \CaHK photometry, as has been done for the main \Pristine survey \citep[][hereafter M24]{martin24}. The XP data has also been used to create a homogeneous all-sky total Galactic extinction (TGE) map \citep{delchambre23}. Correcting for extinction remains a core challenge in the derivation of photometric metallicities, and in this work we derive an empirical calibration for the employed extinction maps by minimising trends between metallicities and E(B$-$V). 

One of the main goals of PIGS is to provide metal-poor candidates for spectroscopic follow-up. We have performed our own low-/medium-resolution follow-up with the Anglo-Australian Telescope (AAT) and the AAOmega+2dF instrument \citep{saunders04, lewis02, sharp06}, resulting in spectra for $\sim13,000$ stars, of which $\sim 7000$ have $\feh < -1.5$ and $\sim 1800$ have $\feh < -2.0$ \citep{arentsen20_II, arentsen24}. We have used these data to fine-tune our AAT follow-up target selection, to study the orbital properties of metal-poor inner Galaxy stars \citep{arentsen20_I, arentsen24}, to study diffuse interstellar bands \citep{zhao23}, to investigate the frequency of carbon-enhanced metal-poor (CEMP) stars in the inner Galaxy \citep{arentsen21} and in the Sagittarius (Sgr) dwarf galaxy \citep{sestito24_sgrcemp}, to calibrate PIGS photometric metallicities for the study of the metallicity distribution of the Sgr dwarf galaxy \citep{vitali22} and to select Sgr stars for spectroscopic follow-up \citep{vitali25}, and finally to select very and extremely metal-poor inner Galaxy and Sgr stars for high-resolution spectroscopic follow-up and detailed chemical analysis \citep[][PIGS team in prep]{sestito23, mashonkina23, sestito24_sgrvmp, salgado26}. The next step for PIGS will be homogeneous spectroscopic follow-up with the low-resolution mode of the 4-metre Multi-Object Spectroscopic Telescope \citep[4MOST][]{4most}, for which we describe the target selection and observing strategy in this paper. 

Since the start of PIGS in 2017, other efforts at building large (spectro)photometric metallicity datasets in the inner Galaxy have appeared as well. One example is from the Blanco DECam Bulge Survey (BDBS) by \citet{johnson22}, focusing on red clump stars, which have relatively high metallicity (with $\feh > -1$). They adopted a linear relation between $(u-i)_0$ and spectroscopic metallicities from the GIRAFFE Inner Bulge Survey \citep[GIBS;][]{zoccali17} to derive photometric metallicities. The metallicities from this survey are not calibrated for metal-poor stars (and maybe not sensitive enough) or for red giant branch stars, and are not publicly available yet. Then the \Gaia DR3 XP spectra allowed for homogeneous all-sky determinations of metallicities across stellar types, such as the \texttt{XGBoost} analysis by \citet{andrae23} trained on APOGEE DR17 \citep{majewski17, apogeedr17} and supplemented with a dedicated very metal-poor sample from \citet{li22}, or the \texttt{StarHorse} \texttt{XGBoost} analysis by \citet{khalatyan24} trained on various major spectroscopic surveys. \citet{rix22} used an earlier version of the \citet{andrae23} dataset combined with \Gaia DR3 radial velocities to reveal a centrally concentrated ``poor old heart'' of the Milky Way. With PIGS, we can produce good photometric metallicities for fainter stars than with \Gaia DR3 XP, which has been key for the very metal-poor inner Galaxy target selection for the 4MOST MIlky way Disk And BuLgE Low-Resolution Survey \citep[4MIDABLE-LR,][]{chiappini19}.  

We describe the observations, data reduction, calibration and release of the PIGS \CaHK photometry in Section~\ref{sec:cahksurvey}. The photometric metallicities, extinction correction, validation against spectroscopic data and recommended quality cuts on the released photometric metallicities are presented in Section~\ref{sec:photmet}. We assess the potential impact of interstellar Ca H\&K lines on our photometric metallicities in Section~\ref{sec:interstellar}. Finally, the target selection and goals of the 4MOST \texttt{VMP\_PIGS} sub-survey are described in Section~\ref{sec:4most}, before we summarise our results in Section~\ref{sec:summary}.

\section{The PIGS \CaHK survey}\label{sec:cahksurvey}

\subsection{Observations}

Each individual CFHT MegaCam \CaHK pointing for PIGS was observed for 200s, the same as for the main \Pristine survey. Observing constraints were set to seeing $<0.8''$ and airmass $<2$. The seeing limit is stricter for PIGS compared to the main \textit{Pristine} survey (\citealt{starkenburg17b}, \citetalias{martin24}), which is necessary due to the crowding in the inner Galaxy. The airmass limit is relatively high because the bulge is low on the sky from the Northern hemisphere and this choice optimised our chances of getting all fields observed with sufficient quality. 

Our strategy resulted in an uncertainty of $\lesssim0.05$ for \CaHK $\sim 21$ in a typical, relatively dense survey field ($l=+4.2^\circ$, $b=-6.6^\circ$). What magnitude this is reached at depends on the colour of the stars. In this work, we use \Gaia ($G_\mathrm{BP} - G_\mathrm{RP}$), hereafter (BP$-$RP). Stars with (BP$-$RP) = $1.0$ and $1.5$ reach this median \CaHK uncertainty for \Gaia $G$ magnitude of $\sim 19$ and $18$, respectively. A median \CaHK uncertainty of 0.1 is reached at $G\sim19.5$ for most stars. Appendix Figure~\ref{fig:footdepth} shows the median \CaHK uncertainty across the PIGS footprint for stars with $16.5 < G < 17$ and $1.3 < \mathrm{BP}-\mathrm{RP} < 1.5$.

\subsection{Footprint}
\begin{figure}
\centering
\includegraphics[width=1.0\hsize,trim={0.0cm 0.0cm 0.0cm 0.0cm}]{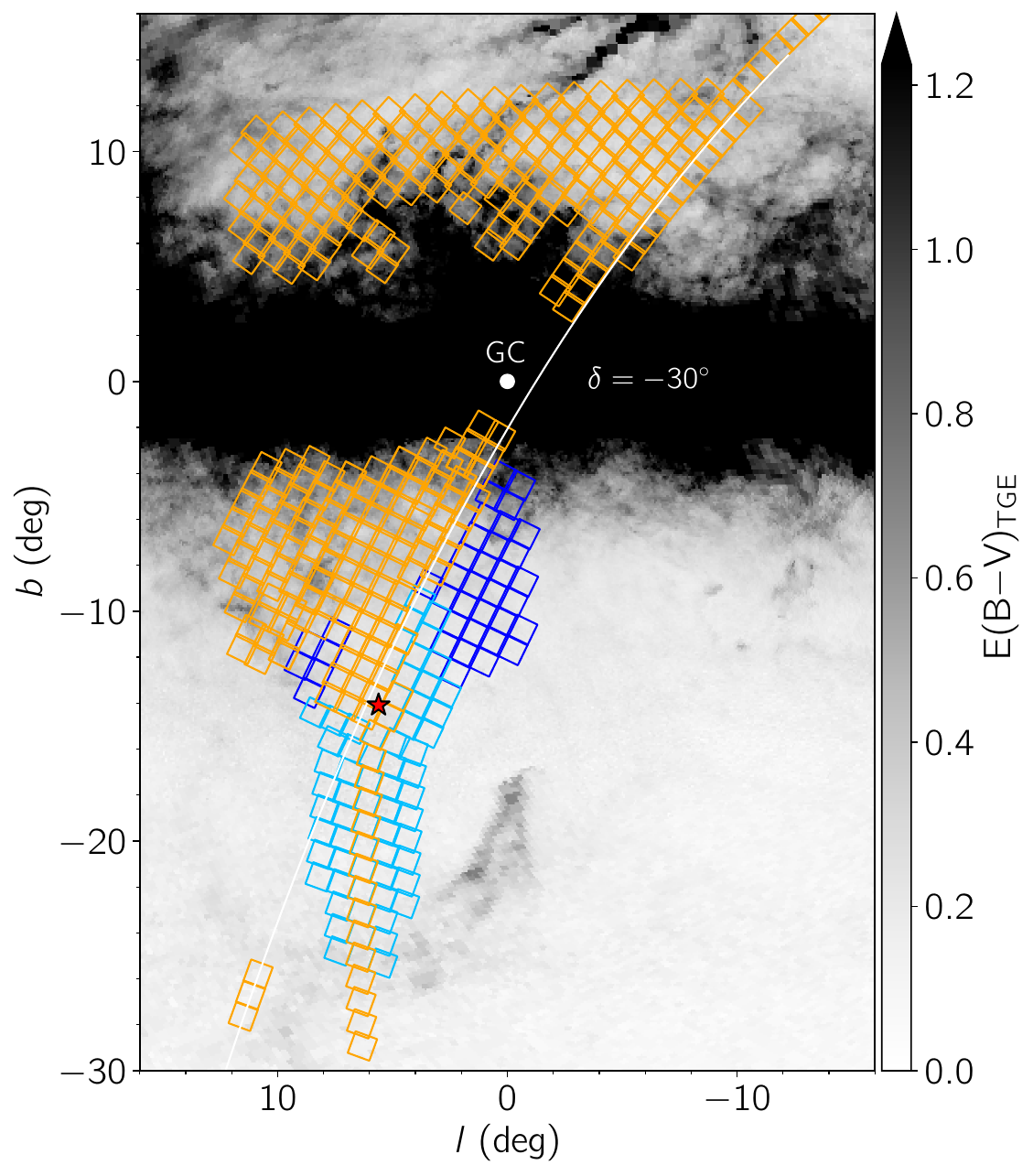}
\caption{PIGS photometric footprint, with \Gaia's TGE extinction map \citep{delchambre23} in the background (where the map is not available near the plane, the colour has been set to the saturation limit). Here, E(B$-$V)$_\mathrm{TGE}$ has been computed as $0.95 \times A_{0,\mathrm{TGE}} / 3.1$.
The orange fields are the main PIGS fields, while blue fields were observed as part of the main \textit{Pristine} survey -- with light fields following the main survey strategy and dark fields the PIGS strategy. M54, at the heart of the Sgr dwarf galaxy, is indicated with a red star marker.
}
\label{fig:footprint} 
\end{figure}

Our main goal was to cover as much of the inner Galaxy with \CaHK photometry as possible. The PIGS footprint is presented in Figure~\ref{fig:footprint}, with each square indicating a CFHT image/field. We avoided regions with very high extinction, since our filter is very blue with the reddening correction going as $\sim 4 \times$ E(B$-$V), and we could not cover one quadrant of the inner Galaxy ($b < 0^{\circ}$ and $l < 0^{\circ}$) due to CFHT being located in the northern hemisphere. We aimed to cover all regions with E(B$-$V) $< 0.7$, $|b| < 12^{\circ}$ and $0^{\circ} < l < 12^{\circ}$, with an extension to $-10^{\circ} \lesssim l < 0^{\circ}$ in the northern half. The overall coverage of the PIGS footprint is $340$~deg$^2$. Most of the fields were part of the dedicated PIGS programme (pilot in 2017 and bulk in 2018-2019, all shown in orange in Figure~\ref{fig:footprint}). 

The rest of the footprint was observed as part of the main \textit{Pristine} survey programme which has been running in snapshot (poor-weather) mode since semester 2016B. The light blue fields were added as filler fields during a period in 2019 where MegaCam was the only active instrument on CFHT. Almost all of these fields were observed in good seeing conditions similar to the rest of PIGS, but with slightly higher sky brightnesses. The variations in depth across the footprint are summarised in Appendix Figure~\ref{fig:footdepth}. The dark blue fields were observed as part of the main \textit{Pristine} survey during 2023, with the goal of completing the footprint for $l > 0^{\circ}$, allowing this region to be included in the 4MOST follow-up. The dark blue fields with $l < 5^{\circ}$ were observed in slightly poorer seeing conditions of $1-2''$. 

We extended the scope of the original inner Galaxy survey to include the Sgr dwarf galaxy, which is the part of the footprint with $-30^{\circ} < b \lesssim -12^{\circ}$, covering roughly $50$~deg$^2$. The centre of Sgr, M54, is located at $(l,b) \sim (5.6^{\circ}, -14.1^{\circ})$. Most of the Sgr fields were observed as part of the main \textit{Pristine} survey, apart from the northernmost region (around M54), a strip down the semi-major axis of Sgr and three off-axis fields around $(l,b) \sim (11^{\circ}, -26^{\circ})$. 

\subsection{Data reduction and calibration}\label{sec:calib}

Here we briefly describe the data reduction and calibration strategy for the PIGS photometry. For the data release in this work, we re-reduce all the photometry homogeneously. We follow the main \textit{Pristine} survey strategy, for which details can be found in \citetalias{martin24} -- here we will only repeat the main relevant aspects. 

After initial processing at CFHT, we refine the astrometry of the images using \Gaia DR2\footnote{DR2 is also used for the main \Pristine survey. \citetalias{martin24} tested that changing to (E)DR3 does not make a significant difference -- the rms on the astrometry is typically better than 0.1'', which is more than enough for our purpose. }. The Cambridge Astronomy Survey Unit (CASU) pipeline \citep{casu} is used on the resulting images to perform aperture photometry. Given that our images are relatively shallow, crowding is not a major issue. This is especially the case for brighter stars (e.g. $G<17$), although the fraction of sources flagged as extended objects (\texttt{CASU\_flag} = +1) does increase with increasing magnitude (e.g. on average $\sim3\%$ at $\CaHK = 16$, $\sim6\%$ at $\CaHK = 18$ and $\sim 20\%$ at $\CaHK = 20$, but it depends where in the footprint). 
Next, the images need to be corrected for variable zero-point offsets, including variations as a function of the location in the field of view (FoV), which can vary between runs (we define a run as each time MegaCam is placed back on the telescope), see \citetalias{martin24}. This is done by comparing the MegaCam \CaHK magnitudes with synthetic \CaHK magnitudes derived from the \Gaia XP spectra \citep{deangeli23} by \citetalias{martin24} (hereafter $CaHK_\mathrm{syn}$). We adopt the FoV models determined for the main \Pristine survey wherever there were PIGS and main survey fields observed in the same MegaCam run, and derive new models with \texttt{PhotCalib}\footnote{\url{https://github.com/zyuan-astro/PhotCalib}} based on the PIGS observations for the runs in 2023. 

The overall zero-point correction for each image is determined from the difference between the MegaCam and synthetic \CaHK magnitudes for a clean sample of stars in each image. Only stars passing the following quality cuts were used in this step: 

\begin{itemize}
        \item no star within 10 half-light radii from the centre of globular clusters (similar to \citetalias{martin24})
        \item \texttt{CASU\_flag} = $-1$ (as in \citetalias{martin24})
        \item $\delta CaHK_\mathrm{syn} < 0.1$ (as in \citetalias{martin24})
        \item $\delta CaHK < 0.05$
        \item $P_\mathrm{var} <0.3$
\end{itemize}

where $P_\mathrm{var}$ is the probability that a star is variable based on its \Gaia G band fractional flux uncertainty times the square root of the number of \Gaia observations (this combined quantity is \texttt{gvar}), as in \citetalias{martin24}. The median difference after iteratively removing $2\sigma$ outliers is adopted as the zero-point correction. For the fields observed in 2023 we follow the exact same approach as in \citetalias{martin24} instead, as they were fully processed with the main survey. 

For the J-PLUS survey, \citet{jplus} performed a similar photometric calibration using synthetic \Gaia XP photometry for their filters, and found that there are systematics in the comparison between their photometry and \Gaia synthetic photometry with magnitude and colour. Our calibration was done before this work came out and does not take this into account. We check what the potential effect of ignoring this might be on our zero-point calibrations in the Appendix Figure~\ref{fig:cal2}, and do not find evidence that this significantly impacts our calibration. We therefore do not implement any recalibration in this work. 

\begin{figure}
\centering
\includegraphics[width=0.7\hsize,trim={0.0cm 0.4cm 0.0cm 0.0cm}]{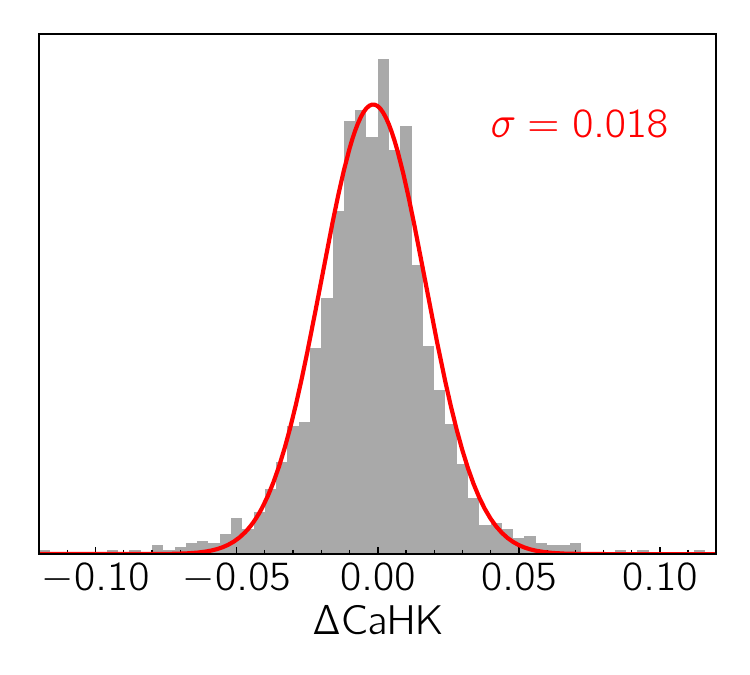}
\caption{Difference in calibrated \CaHK magnitudes between duplicate stars in different fields for run 17Am07. A Gaussian with $\mu = 0.001, \sigma = 0.018$ mag (the median and dispersion of the distribution) is over-plotted. Only stars with \CaHK measurement uncertainties $< 0.01$ were used in this exercise.}
\label{fig:calib} 
\end{figure}

After applying the FoV and zero-point corrections, we arrive at the calibrated \CaHK magnitudes for each field. The resulting distribution of the difference between calibrated \CaHK magnitudes for duplicate stars in overlapping fields for one of the runs (17Am07) is show in Figure~\ref{fig:calib}. Largely by construction (but encouragingly) the median difference is zero, with a dispersion of $\sim 18$ mmag resulting in an uncertainty floor of $18/\sqrt{2} = 13$ mmag -- similar to what was found for the main \Pristine survey (\citetalias{martin24}). 

We note that the target selection for the spectroscopic follow-up presented in previous PIGS papers has been based on older versions of the catalogue from different versions of the data reduction and calibration that have evolved as the survey went along. The zero points for the PIGS images in the inner Galaxy fields were previously determined in a relative manner using the same stars in an overlapping region with a neighbouring field, with one field being chosen as a reference \citep[see][]{arentsen20_II}. The FoV correction was the old version as described in \citet{starkenburg17b}, the same for all runs. The relative approach worked well to locally select the most metal-poor stars, as evidenced by the successful PIGS follow-up campaign, but created some trends across the wider footprint and was only calibrated internally. For the PIGS Sgr fields, a similar approach to \citet{starkenburg17b} was taken instead for the zero-point normalisation, using the stellar locus but fitting for extinction simultaneously  \citep{vitali22}. Now that synthetic \CaHK magnitudes are available thanks to the Gaia DR3 XP spectra, everything can be calibrated homogeneously, allowing all of PIGS to be placed on the same scale as the main \Pristine survey.

\subsection{\CaHK catalogue}

A significant fraction of stars have multiple detections in PIGS, because of intentional $\sim0.1^\circ$ overlaps between different pointings due to our original calibration strategy, as well as thanks to the ``ears'' of the MegaCam detector\footnote{\url{https://www.cfht.hawaii.edu/Instruments/Imaging/Megacam/specsinformation.html}}. We merge these as in \citetalias{martin24} to improve their signal-to-noise and update the \texttt{CASU\_flag} accordingly into a new \texttt{merged\_CASU\_flag}. The merged\_CASU\_flag is $-9$ for saturated stars, $-2$ for likely point sources, $-1$ for very likely point sources, $+1$ for extended sources, $+2$ for merged sources with discrepant flags from individual detections and $+3$ for merged sources with discrepant photometry at the $>5\sigma$ level. 

We publish the calibrated merged PIGS \CaHK catalogue with this paper, for all 55 million sources with a cross-match in \Gaia DR3 and \CaHK uncertainties less than 0.3~mag -- see the first nine columns of Table~\ref{tab:photmet_columns}. Recommended quality cuts on the photometric metallicities are described in Section~\ref{sec:quality}, which largely also apply to the use of \CaHK photometry in general. 

\section{Photometric metallicities}\label{sec:photmet}

We use the same model as in \citetalias{martin24} with minor modifications\footnote{The code and the training sample are publicly available at \url{https://github.com/ankearentsen/Pristine_code}. Note that in this work we use and release results from the \Pristine DR2 version of the code. The 4MOST/4MIDABLE-LR catalogue, submitted in 2023, was prepared with the DR1 version, but as described in the \Pristine DR2 paper \citep{yuan26}, the differences are negligible for the vast majority of objects.} 
to derive photometric metallicities from the PIGS \CaHK and \Gaia DR3 G, BP and RP broad-band photometry, with details described in that work and in \citet{starkenburg17b}. In short, a spectroscopic training sample was used to build a model of the metallicity for each position in the (BP$-$RP)$_0$ vs. (\CaHK~$-$~G)$_0 -2.5$(BP$-$RP)$_0$ colour-colour space. The colour-colour space was divided into a grid, then the mean metallicity of the training sample was determined in each grid cell and finally the entire space was smoothed by interpolation between the cells. Models were built separately for giants ($\logg < 3.9$) and dwarfs ($\logg > 3.9$ or $\teff > 5800$~K). \emph{In this work we will only use the giant model.} For each star, based on their position in the colour-colour space and the uncertainties in the photometry, the code provides values of the 16th, 50th and 84th percentiles for the metallicity derived from 100 Monte Carlo samples, as well as \texttt{mcfrac}, the fraction of samples for each star that falls within the limits of the metallicity model. The metallicity limits are \feh = [$-4.0, 0.0$]. The emphasis in the \Pristine photometric metallicity training sample and model building was on metal-poor stars, so the results are most appropriate for stars with $\feh < -1.0$. Stars exactly at $\feh = -4.0$ should generally be excluded, and stars with $\feh < -3.5$ require stricter quality cuts (see Section~\ref{sec:photmetcat}) and/or further verification depending on the science case. 

\begin{figure}
\centering
\includegraphics[width=1.0\hsize,trim={0.0cm 1.0cm 0.0cm 0.0cm}]{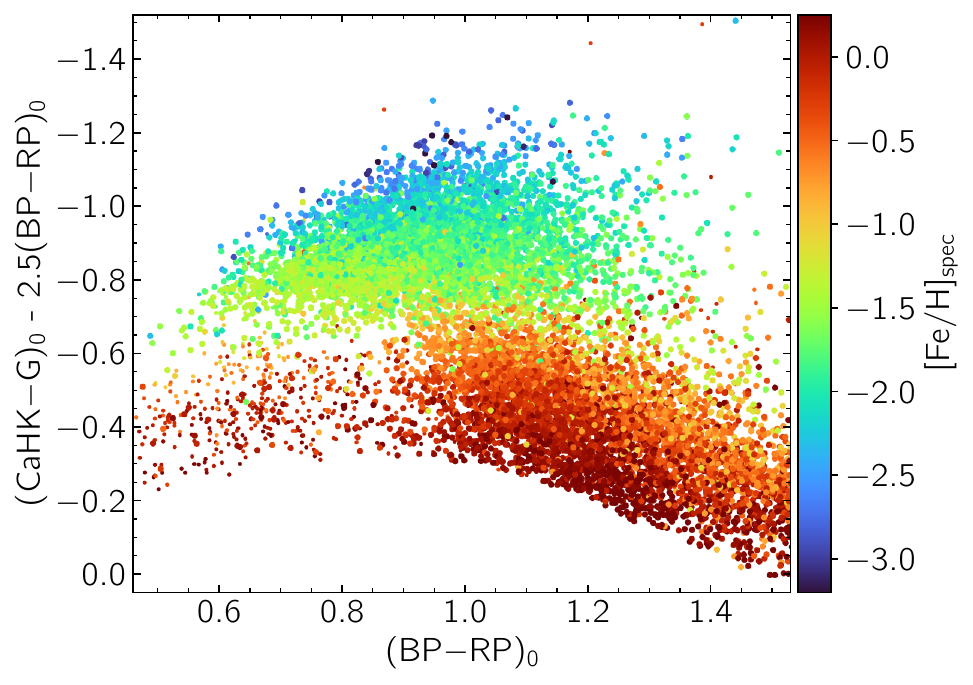}
\caption{\Pristine colour-colour diagram for stars in the PIGS footprint with known spectroscopic metallicities (from APOGEE, PIGS/AAT and high-resolution follow-up, see Section~\ref{sec:spec}) that pass all quality cuts in Section~\ref{sec:quality}, apart from the colour and parallax cuts. Dwarf stars ($\logg > 3.8$) are shown in smaller points. The extent of the x-axis at the red end is set by the limit of the \Pristine training grid, within which the dereddening is performed. Sgr dwarf galaxy stars have been excluded. 
}
\label{fig:ccd} 
\end{figure}

Figure~\ref{fig:ccd} presents the \Pristine colour-colour space used in the metallicity code, populated with stars with known spectroscopic metallicities from various sources and using the PIGS \CaHK photometry. The extinction has been corrected for as described in Section~\ref{sec:ext}, and the quality cuts recommended in Section~\ref{sec:quality} have been applied. Only stars within the valid colour range for the photometric metallicity code are shown (limited to (BP$-$RP)$_0 < 1.5$). 

Throughout this work, we refer to the photometric metallicity as \fehphot. When necessary to distinguish between results using different extinction maps, they are described as [Fe/H]$_\mathrm{phot,TGE}$ or [Fe/H]$_\mathrm{phot,bay}$. In Section~\ref{sec:spec}, we identify a 0.35~dex offset between our photometric metallicities and spectroscopic metallicities from the literature, and subtract this to derive [Fe/H]$_\mathrm{phot,TGE,corr}$. The latter is only used for target selection and when counting numbers of stars at low metallicity, is never used in any figures. 

\subsection{Extinction correction}\label{sec:ext}

An important difference between PIGS and the main \Pristine survey is that we are dealing with stars in high extinction regions. The \citet[][hereafter SFD]{schlegel98} extinction map used in \citetalias{martin24} is not appropriate for the inner Galaxy, so other options have to be explored. For previous versions of the PIGS photometric catalogues used for target selection the \citet{green19} Bayestar map was adopted, see \citet{arentsen20_II}. We use it again in this work, but since it is not available for the full footprint (regions with $\delta \lesssim -30^\circ$), we also use the \Gaia DR3 2D Total Galactic Extinction map \citep[TGE,][]{delchambre23}.

We note that given our footprint and our sample of interest of distant inner Galaxy stars ($d \gtrsim 4$~kpc), a 3D map is not ultimately necessary. Most MW dust is located within $\sim 0.1-0.2$~kpc of the Galactic plane \citep{drimmel01}, which at an angle of $5$~deg above or below the plane is reached after $\sim 1-2.5$~kpc. As stated in Section~\ref{sec:quality}, we attempt to remove nearby stars when estimating photometric metallicities.

\begin{figure}
\centering
\includegraphics[width=0.9\hsize,trim={0.0cm 0.2cm 0.0cm 0.0cm}]{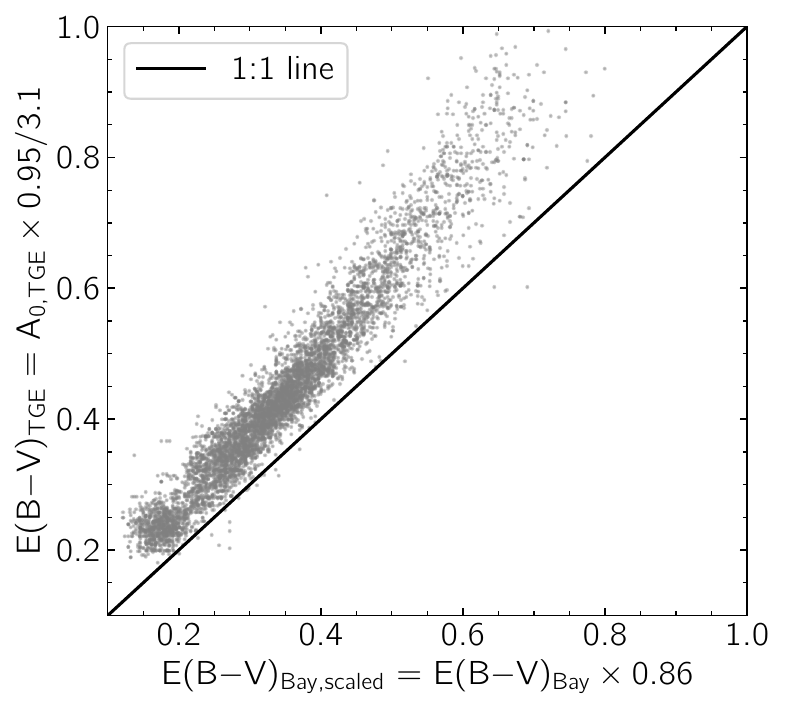}
\caption{Comparison of E(B$-$V) from the Bayestar and TGE maps for stars with PIGS/AAT spectra. The 0.86 factor on the x-axis is the SFD map rescaling factor determined by \citet{schlafly11}, the 3.1 and 0.95 factors on the y-axis are the assumed $R_V$ and an estimate of A$_V$/A$_0$ for giants, respectively.}
\label{fig:AVcomp} 
\end{figure}

The Bayestar map is a three-dimensional map that has been built using PanSTARRS \citep{panstarrs} and 2MASS \citep{2mass} photometry plus \Gaia distance information for individual stars. The assumed extinction law is from \citet{schlafly16} and the reddening is provided in a unit similar to that of the SFD map. This is not exactly the same as E(B$-$V), as \citet{schlafly11} found that the SFD map somewhat overestimates E(B$-$V) and should be multiplied by a factor of $\sim 0.86$. 
We do this for the Bayestar map as well, since its reddening unit is defined to be on a similar scale to SFD \citep{green19}, when converting its reddening to extinction in specific filters using the \citetalias{martin24} extinction coefficients. 
We set the distance equal to 8~kpc for all stars when we query the Bayestar map, but as stated above, most dust will be in the foreground so the precise input distance does not matter if we are interested in inner Galaxy stars.  

\begin{figure*}
\centering
\includegraphics[width=0.9\hsize,trim={0.0cm 0.0cm 0.0cm 0.0cm}]{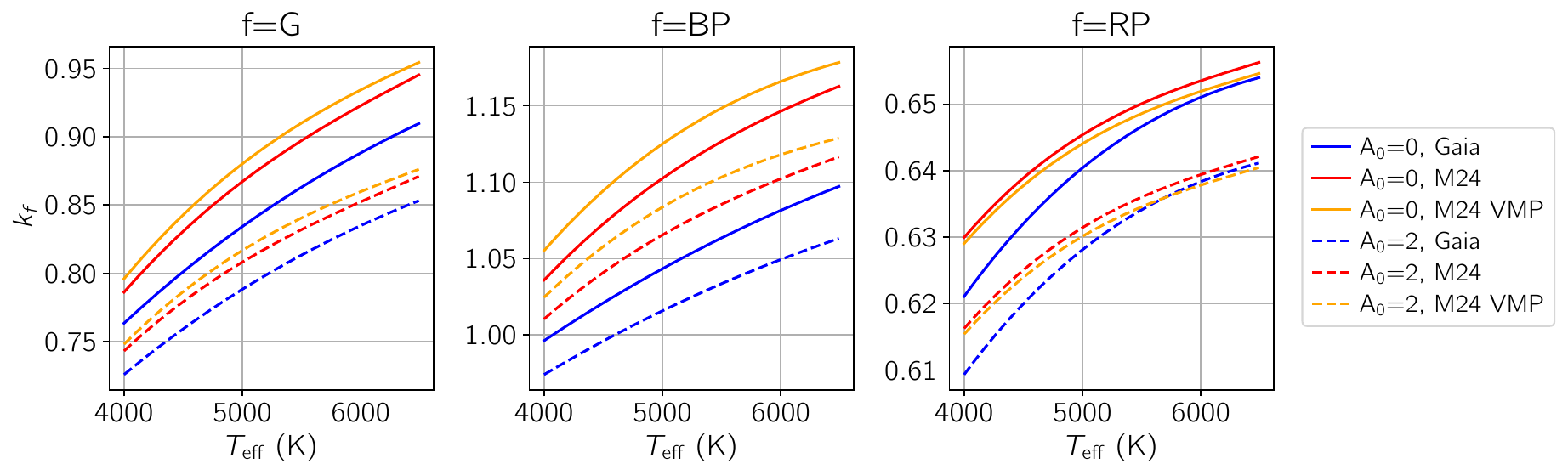}
\caption{Comparison of \Gaia (blue) and \Pristine \citetalias{martin24} (red) extinction coefficients for the \Gaia filters, both for giant stars. The \citetalias{martin24} coefficients are shown for both $\feh = 0$, to match the assumption for \Gaia (red), and $\feh = -2.0$ (orange), to show the dependence on metallicity.}
\label{fig:coeff} 
\end{figure*}

The TGE map is a two-dimensional map built using \Gaia DR3 \texttt{Apsis} pipeline \citep{creevey23} products for individual stars, and approximates the total extinction along the line of sight. The assumed extinction law is \citet{fitzpatrick99} and the extinction is provided in units of A$_0$ (the monochromatic extinction at 541.4 nm), similar to A$_V$ but not exactly the same due to the slightly different central wavelength and width of the $V$ filter. 

We compare the extinction values for the Bayestar and TGE maps for the PIGS spectroscopic dataset in Figure~\ref{fig:AVcomp}, converting both to E(B$-$V)\footnote{We note that while here in this figure we use a constant factor 0.95 to convert A$_0$ to A$_V$, in the \Pristine photometric metallicity code this factor depends on \teff, A$_0$ and \feh, as for the \Gaia filters, and is determined iteratively.}. There is a significant offset between the two (median = 0.09), as well as a trend. This discrepancy was discussed to some degree in the \Gaia GSP-Phot paper \citep{andrae23_gaia}, but not investigated in detail. This difference has a significant impact on the current work, where we need to deredden photometry in multiple bands in a high-extinction region to derive photometric metallicities from a colour-colour space. 

\subsubsection{Extinction coefficients}

Another challenge in deriving the extinction corrections is that the \Gaia photometric bands are very broad, which makes the extinction coefficient (the factor to multiply with E(B$-$V) or A$_0$) dependent on the spectral energy distribution of each star -- affected by the effective temperature and the reddening itself, as well as a minor contribution from metallicity. This was addressed in \citetalias{martin24} for the \Pristine photometric metallicity code by building a polynomial model for the extinction coefficients as a function of \teff, $A_0$ and \feh, adopting the \citet{fitzpatrick99} extinction law (with $R_V = 3.1$). The extinction coefficients are implemented inside the photometric metallicity estimation procedure with the extinction coefficients and the photometric \teff and \feh updating over a few iterations. 

The \citetalias{martin24} extinction coefficients were computed with an early version of the \texttt{dustapprox} package \citep{dustapprox}, which, as we discovered during this work, applies the extinction curve in magnitudes as an optical depth ($e^{-A_\lambda}$ instead of $10^{-0.4A_\lambda}$). This makes the coefficients per unit $A_0$ a factor $2.5/\ln 10 \approx 1.086$ too large. The normalisation does not affect our photometric metallicity estimates derived with E(B$-$V)-based maps (SFD, Bayestar), since it cancels in the conversion from E(B$-$V) to $A_0$, and for TGE it is absorbed by the empirical scaling factor derived below.

The adopted coefficients are slightly different to the published Gaia polynomial\footnote{\url{https://www.cosmos.esa.int/web/gaia/edr3-extinction-law}}, which was derived using a fixed, solar metallicity and assuming the \citet{fitzpatrick19} extinction law (according to the webpage). The comparison between the \Gaia and \citetalias{martin24} relations is shown in Figure~\ref{fig:coeff}. Given the otherwise very similar approach in deriving the two sets of coefficients, this is mainly due to the issue in \texttt{dustapprox}, together with the different reference wavelength of $A_0$ (541.4\,nm here and 550\,nm for the \Gaia coefficients) and the different assumed extinction laws. 
We adopt the \citetalias{martin24} coefficients because they include a dependence on metallicity and adopt the same extinction law as TGE, but we also test the \Gaia coefficients. 

\subsubsection{Recalibrating the extinction}\label{sec:extcal}

After our first derivation of the photometric metallicities, we noticed strong trends with E(B$-$V) in the difference between photometric and spectroscopic metallicities when using the TGE map. There should not be any trends in the \textit{spectroscopic} metallicities with extinction, and none have been seen previously in PIGS, so fault must be with the photometric metallicities. 

We used the \citetalias{martin24} extinction coefficients, and tested whether this trend disappeared when adopting the \Gaia coefficients, but it did not -- it biased them in the opposite direction, with a similar magnitude. We later realised that this was largely because the \CaHK coefficients and the \Gaia coefficients have a different normalisation (see the \texttt{dustapprox} issue above). Putting them on the same scale, the \Gaia coefficients show a similar trend and overall metallicity offset.

\begin{figure}
\centering
\includegraphics[width=0.9\hsize,trim={0.0cm 0.5cm 0.0cm 0.0cm}]{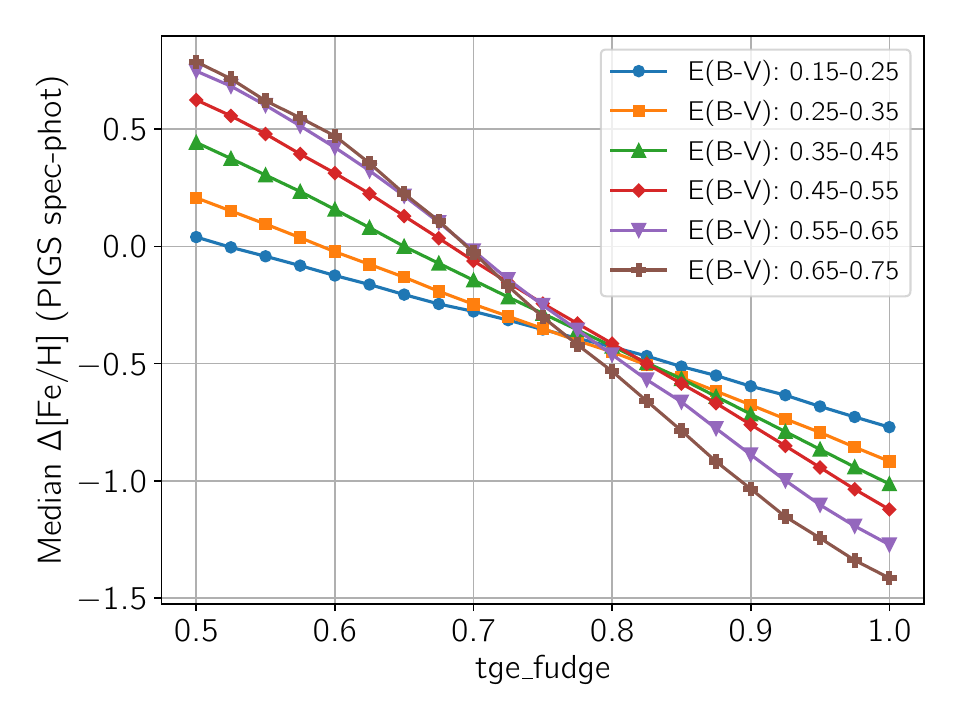}
\includegraphics[width=0.9\hsize,trim={0.0cm 0.5cm 0.0cm 0.0cm}]{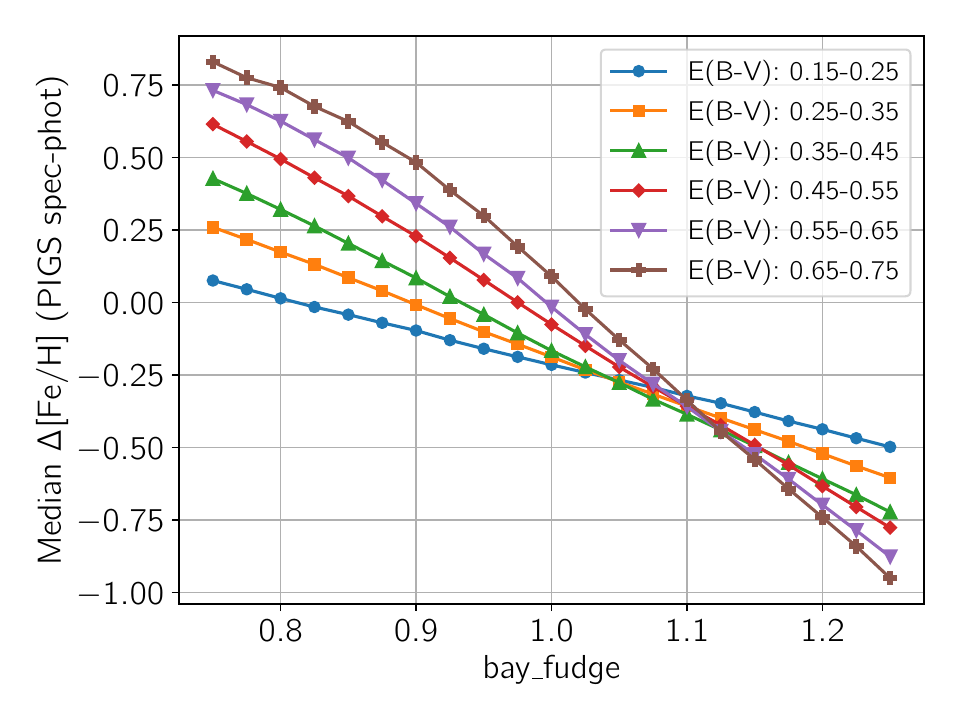}
\caption{Median difference in photometric versus spectroscopic metallicity as a function of the fudge factor, in different bins of E(B$-$V) from the respective extinction maps. Where these lines converge is the optimal choice for the fudge factor to reduce the metallicity-extinction bias. Top: \texttt{tge\_fudge} for TGE, with 0.76 the optimal value, bottom: \texttt{bay\_fudge} for Bayestar, with 1.1 the optimal value.}
\label{fig:fudge} 
\end{figure}

We then tested whether multiplying the TGE extinction values by a constant factor could alleviate these systematic differences with extinction. We re-ran the photometric metallicities for the PIGS spectroscopic sample \citep{arentsen24} with a fudge factor \texttt{tge\_fudge} in the range from 0.5 to 1.0 with steps of 0.025. We apply all quality cuts described in the next section and select only metal-poor giant stars with $-2.5 < \feh < -1.5$ and $\logg < 3.5$ according to the \FERRE analysis of the PIGS spectra \citep{arentsen20_II}. Figure~\ref{fig:fudge} shows the result of this test, presenting the spectroscopic vs. photometric metallicity difference as a function of this factor in bins of E(B$-$V). The difference between extinction bins is minimal around \texttt{tge\_fudge} $\sim 0.75$. After refining the \texttt{tge\_fudge} range, we settle on a scaling factor of 0.76. We note that for this \texttt{tge\_fudge}, there is a systematic offset between the photometric and spectroscopic metallicities of $\sim 0.35$~dex. We will further discuss this in the next section. 

When using the Bayestar map, we did not notice as obvious a trend with extinction for the metallicity difference, but given the results above we decided to do the same test to check. The results are shown in the bottom panel of Figure~\ref{fig:fudge}, and suggest that a \texttt{bay\_fudge} could indeed reduce a systematic shift in metallicity with extinction, with an optimal value around 1.1. 
Since we originally scaled the Bayestar map reddening values down by 0.86, this results in an overall scaling of $1.1\times0.86 = 0.946$. This is consistent with the expected conversion of the Bayestar19 unit to E(B$-$V) for cool giants ($\sim$0.95), so the 0.86 factor appears to be too small for Bayestar19 in practice rather than the map requiring a correction.
We apply this scaling throughout this work. Interestingly, we note that the systematic offset between the spectroscopic and photometric metallicities for this optimal \texttt{bay\_fudge} is the same as for the optimal TGE case above, namely $\sim 0.35$~dex. 

\subsubsection{Spatial extinction variations}

To test whether there are systematic differences in the extinction maps between positive and negative latitudes, we redo the analysis above but split into $b<0^\circ$ and $b>0^\circ$. While the samples get smaller per extinction bin, we do find that for both the TGE and Bayestar maps, the fudge factor is slightly higher for positive latitudes compared to negative latitudes, by $\sim 0.05$. This could potentially be due to variations in the extinction law, but it is beyond the purpose of this paper to pursue this in detail. We note that this could point to potential spatial systematics in our derived photometric metallicities.  

\subsubsection{Summary}

To reduce biases in the photometric metallicities between low and high extinction regions, we adopt  \texttt{tge\_fudge}~$= 0.76$ and \texttt{bay\_fudge}~$= 1.1$ as multiplicative factors for A$_0$ and E(B$-$V) for the TGE and Bayestar maps, respectively. 
These are empirical corrections within the \Pristine photometric metallicity framework, using specific spectroscopic datasets, and are not to be taken as universal scaling factors of the TGE and Bayestar maps (e.g. in contrast with the $\sim 0.86$ rescaling factor for the SFD map by \citealt{schlafly11}). 
For Bayestar, the net factor of 0.946 is consistent with the expected conversion of the Bayestar19 reddening unit to E(B$-$V) for cool giants, whereas for TGE, after correcting for the \texttt{dustapprox} normalisation of the extinction coefficients, \texttt{tge\_fudge}~$=0.76$ corresponds to $\approx 0.83\,A_{0,\mathrm{TGE}}$, implying that the map overestimates the extinction in the PIGS fields by $\sim$20 per cent.

While this was not the original purpose of this work, we find that using photometric metallicity methods might be a useful tool for constraining extinction models, when comparing to spectroscopic metallicities that are independent of extinction. Further work is necessary to test and exploit this in more detail. 

\subsection{Quality control/cuts}\label{sec:quality}

Ultimately, it is up to the user what quality cuts best fit their science case, in terms of completeness and/or purity.\footnote{The authors of this work are happy to discuss and advise.} We recommend applying at least the following basic quality cuts:

\begin{itemize}
    \item |\texttt{merged\_CASU\_flag}| $= 1$ or $2$
    \item \texttt{merged\_d\_CaHK} $<0.1$
    \item PIGS-Gaia cross-match \texttt{angDist} $< 0.4''$
    \item $P_\mathrm{var} < 0.7$
    \item $C^* < 5 \times \sigma_C$ 
    \item $\fehphot > -4$ and $\fehphot < 0$
    \item $\delta \fehphot < 0.5$
    \item \texttt{mcfrac} $ > 0.8$
    \item E(B$-$V) $< 0.8$
    \item some selection of giant stars
\end{itemize}

where $C^*$ and $\sigma_C$ are the corrected flux excess and the normalized standard deviation of $C^*$ for the G magnitude of a source \citep[following][]{riello21}. For example cuts on $P_\mathrm{var}$, $\delta \fehphot$ and $C^*$ should be made stricter if a pure sample is more important than sample size/completeness. 

A stricter \texttt{merged\_CASU\_flag} would be $-1$ and $-2$ only. Many of the \texttt{merged\_CASU\_flag} $= +2$ sources are expected to be reliable (see Section~\ref{sec:apopigs} for a limited comparison for APOGEE), e.g. if they are combinations of flags $-1$ and $-2$, but others will not -- and the input information is not saved in the merged catalogue. Not including sources with \texttt{merged\_CASU\_flag} = $+2$ reduces the number density of stars in regions that were combinations of multiple observations (re-observed fields, sources in overlaps between fields) but will have lower contamination. When including $+2$ stars, the number of targets goes up by a few percent and there will be a relative overdensity of stars in the merged regions, with some increased contamination. Sources with $+1$ (extended sources, should not usually be present in the \Gaia catalogue but could potentially be chance alignments) could be included too and represent a few percent as well for $G<17$ (for fainter stars, sources with $+1$ and $+2$ together are $\sim$30 per cent of those passing the other cuts). 

Note that the cut on photometric metallicity uncertainty can result in a bias against the lowest metallicity stars at fainter magnitudes. However, significant contamination can come in for these metallicities so making a stricter cut can be necessary (see below).

Finally, the giant star selection is important for two reasons: 1) we have used the \Pristine photometric metallicity model for giants, which is less appropriate for dwarfs (metal-rich, cool dwarfs may masquerade as metal-poor giants), and 2) dwarfs are generally in the foreground when looking towards the inner Galaxy and the assumption that all reddening is in front of these objects is false.

\subsubsection{4MOST target selection}

For the 4MOST target selection (see Section~\ref{sec:4most}) we apply stricter cuts -- those above, with the following changes and additions: 

\begin{itemize}
    \item G $< 17$ 
    \item \texttt{parallax} $-$ \texttt{parallax\_error} $< 0.25$~mas
    \item $0.65 <$ (BP$-$RP)$_0 < 1.3$
    \item $P_\mathrm{var} < 0.7$ or $(\texttt{gvar} < 0.01$ \& G $< 15)$
    \item \texttt{merged\_CASU\_flag} $= -1$ or $-2$
\end{itemize}

The magnitude cut is mostly because of the follow-up strategy (the available observing time and the desired signal-to-noise), and partly due to limitations for fainter objects (contamination from foreground stars, crowding). The parallax cut is to remove most foreground (dwarf) stars and focus on giants in the bulge region, which worked well for our previous PIGS/AAT follow-up \citep{arentsen24}: $95\%$ of fields had $\leq6\%$ dwarf stars ($\logg > 4$) after the \Gaia DR3 parallax cut above. Two fields initially had no parallax cut, their dwarf contamination is reduced from $29\%$ to $6\%$ by making the cut. It becomes less efficient for stars fainter than our previous follow-up ($G > 16.5$) due to the parallax precision, but this will at least somewhat improve with \Gaia DR4. The colour cut focuses on the colour region where the photometric metallicity method works optimally, and where contamination is reduced. Contamination for (BP$-$RP)$_0 > 1.3$ increases due to the fact that giant and dwarf sequences are starting to overlap in this area, as well as an imperfect extinction correction, which can move metal-rich stars into this regime. Contamination for (BP$-$RP)$_0 < 0.65$ rises due to significant numbers of foreground turn-off stars. 
The cut on variability has been loosened for the brightest stars, because the number of stars is smaller so any contamination has a small impact on the overall sample, and even if they are significantly variable given the precise measurements, the magnitude of the variability is unlikely to impact the photometric metallicity. Sources with \texttt{merged\_CASU\_flag} = $+1$ and $+2$ are excluded. Including sources with $+2$ would increase the sample by $\sim 8\%$, but there are a few merged MegaCam fields with spuriously high densities compared to other fields after including them, leading us to make a more conservative selection. 

\subsubsection{Extremely metal-poor stars}\label{sec:emp}

Extremely metal-poor stars (EMP, $\fehphot < -3.0$) are rare, making selections of them sensitive to metal-rich contamination, and lie towards the edge of the photometric metallicity grid, increasing the chance of spurious metallicities. 
Investigating the properties EMP stars, we find that a relative overdensity of $\feh < -3.5$ stars exists predominantly at the fainter end ($G<16$), at higher extinctions, for stars with larger $C^*/\sigma_C$ and $P_\mathrm{var}$ and at redder colours. A selection of EMP stars from our photometry needs much stricter cuts than the minimal recommended ones, such as:
\begin{itemize}
    \item E(B$-$V)~$<0.4$
    \item $C^*/\sigma_C < 1$
    \item $P_\mathrm{var} < 0.3$
    \item $0.65 <$ (BP$-$RP)$_0 < 1.2$
    \item $\fehphot > -3.8$
\end{itemize}
Additional cuts can further improve the purity of the selection: 
\begin{itemize}
    \item \texttt{merged\_CASU\_flag} $= -1$ or $-2$
    \item \texttt{mcfrac} $ > 0.95$
    \item $\delta \fehphot < 0.3$
    \item \texttt{flag\_good\_TGE} $ =1$ and \texttt{flag\_good\_bay} $=1$
\end{itemize}

\subsection{Photometric metallicities catalogue}\label{sec:photmetcat}

\begin{figure}
\centering
\includegraphics[width=0.9\hsize,trim={0.0cm 0.0cm 0.0cm 0.0cm}]{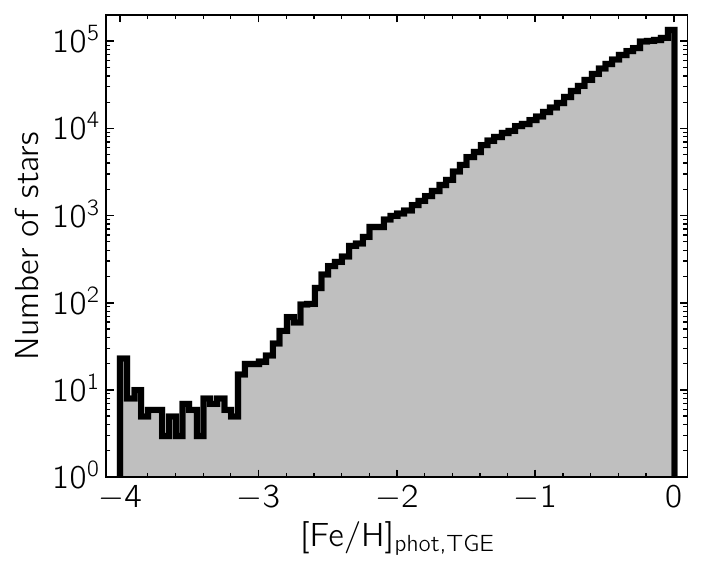}
\caption{Distribution of PIGS photometric metallicities for stars passing all the quality criteria in Section~\ref{sec:quality}.  The metallicities have not been corrected for the identified offset with respect to spectroscopy for stars with $-3.0 \lesssim \feh \lesssim -0.5$ (Section~\ref{sec:spec}), which would shift the distribution $\sim0.35$~dex towards lower metallicities. Note that the limits of the photometric metallicity grid are $\feh = -4.0$ and $0.0$, and that the rising number of objects with $\feh < -3.5$ is likely spurious (see text).
}
\label{fig:mdf} 
\end{figure}

We release TGE and Bayestar photometric metallicities for the $\sim$36 million of the 55 million PIGS stars for which they could be derived,
see Table~\ref{tab:photmet_columns}. Note that the Bayestar metallicities are only available within the PanSTARRS footprint ($\delta \gtrsim -30^\circ$). We include the columns \texttt{flag\_good\_TGE} and \texttt{flag\_good\_bay}, which are 1 for stars that pass the basic quality cuts described in the previous sub-section for the respective extinction map (apart from the giant selection, which is left to the user). \emph{We recommend caution when using the catalogue for stars with $G>17$}, because of the lack of spectroscopic data to test the quality of our photometric metallicities in this regime, and due to increased contamination from foreground (dwarf) populations and increased crowding for fainter stars.

The histogram of photometric metallicities for the 1.2 million stars passing the stricter quality criteria for the 4MOST selection above is shown in Figure~\ref{fig:mdf}. Of those stars, after correcting for the $0.35$~dex metallicity offset discussed in the next section, $254,981 / 17,453 / 519$ have $\feh_\mathrm{phot,TGE,corr} < -1 / -2/ -3$, respectively, corresponding to $20 \%/ 1.4 \%/ 0.04 \%$. Note that these numbers and the Figure are not necessarily representative of the underlying metallicity distribution function of the inner Galaxy, due to the various quality cuts and the mix of stellar populations along the line of sight. 

The metallicity histogram shows a rising number of stars for $\feh < -3.5$, which is not expected. The metallicity grid edge is at $-4.0$, and stars close to the edge are less reliable, especially in this incredibly rare regime where contamination can play a significant role and uncertainties are large. See Section~\ref{sec:emp} for further recommendations.

The typical photometric metallicity uncertainty due to uncertainties in the magnitudes is 0.1~dex for metal-poor stars in this release passing the stricter quality criteria. Note that this uncertainty estimate does not take into account imperfect extinction corrections or limitations of the photometric metallicity model. In the comparison with spectroscopy (see Table~\ref{tab:diff}), the dispersion increases from $\sim 0.2$~dex for E(B$-$V) $< 0.35$ to $\sim0.3$~dex for $0.3 <$ E(B$-$V) $< 0.6$ and $\sim0.45$~dex for $0.6 <$ E(B$-$V) $<0.8$, clearly indicating that the extinction corrections are still suboptimal despite all efforts.

\subsection{Comparisons with inner Galaxy spectroscopic samples}\label{sec:spec}

\begin{figure*}
\centering
\includegraphics[width=0.8\hsize,trim={0.0cm 0.0cm 0.0cm 0.0cm}]{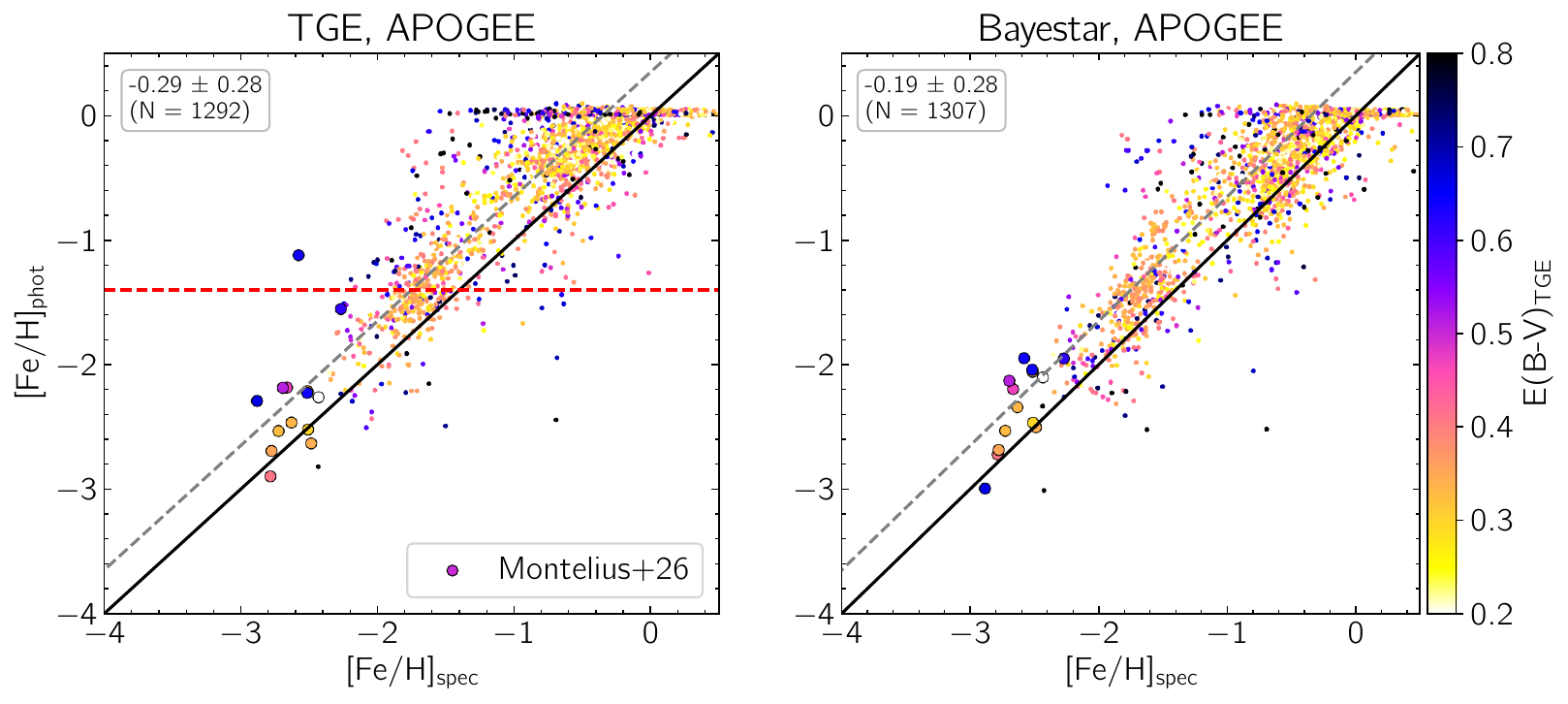}
\includegraphics[width=0.8\hsize,trim={0.0cm 0.0cm 0.0cm 0.0cm}]{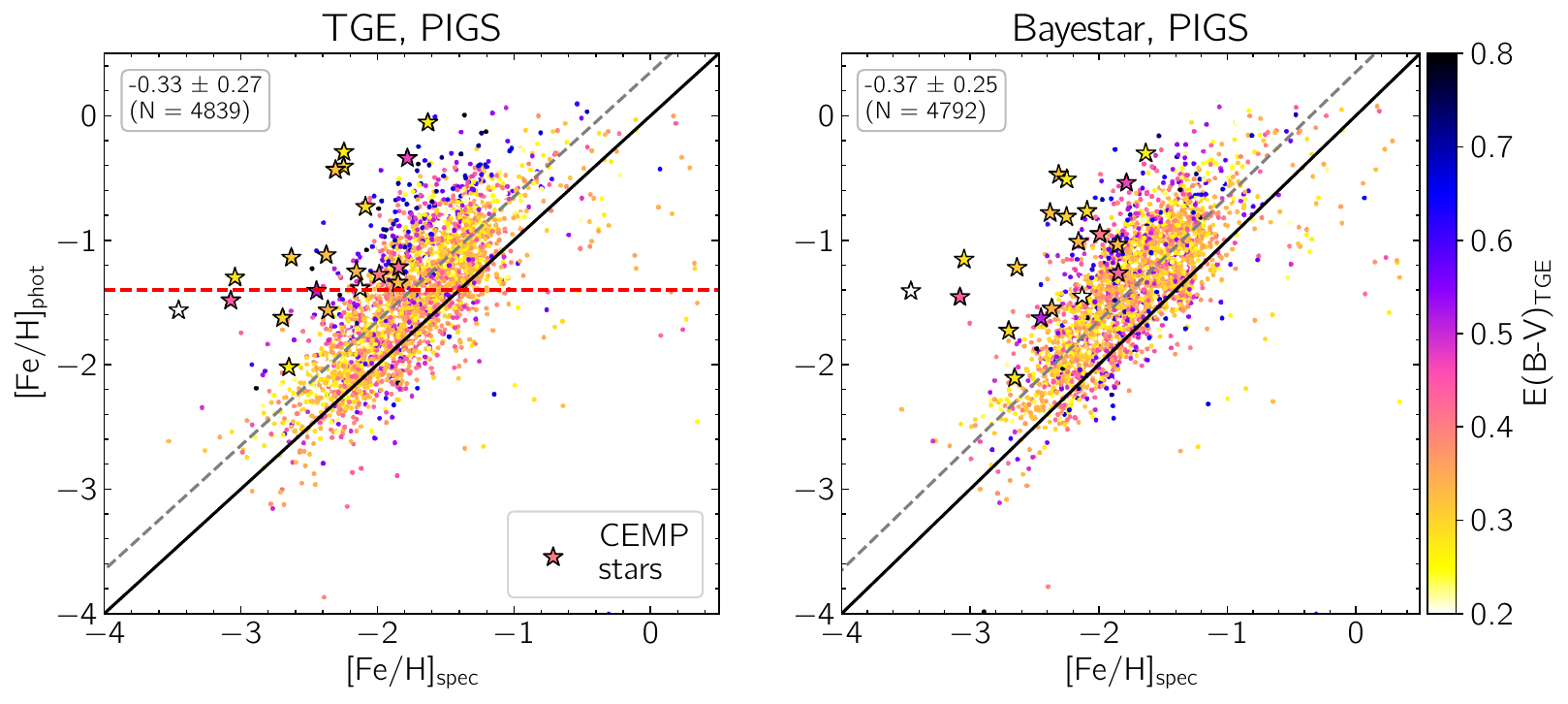}
\includegraphics[width=0.8\hsize,trim={0.0cm 0.0cm 0.0cm 0.0cm}]{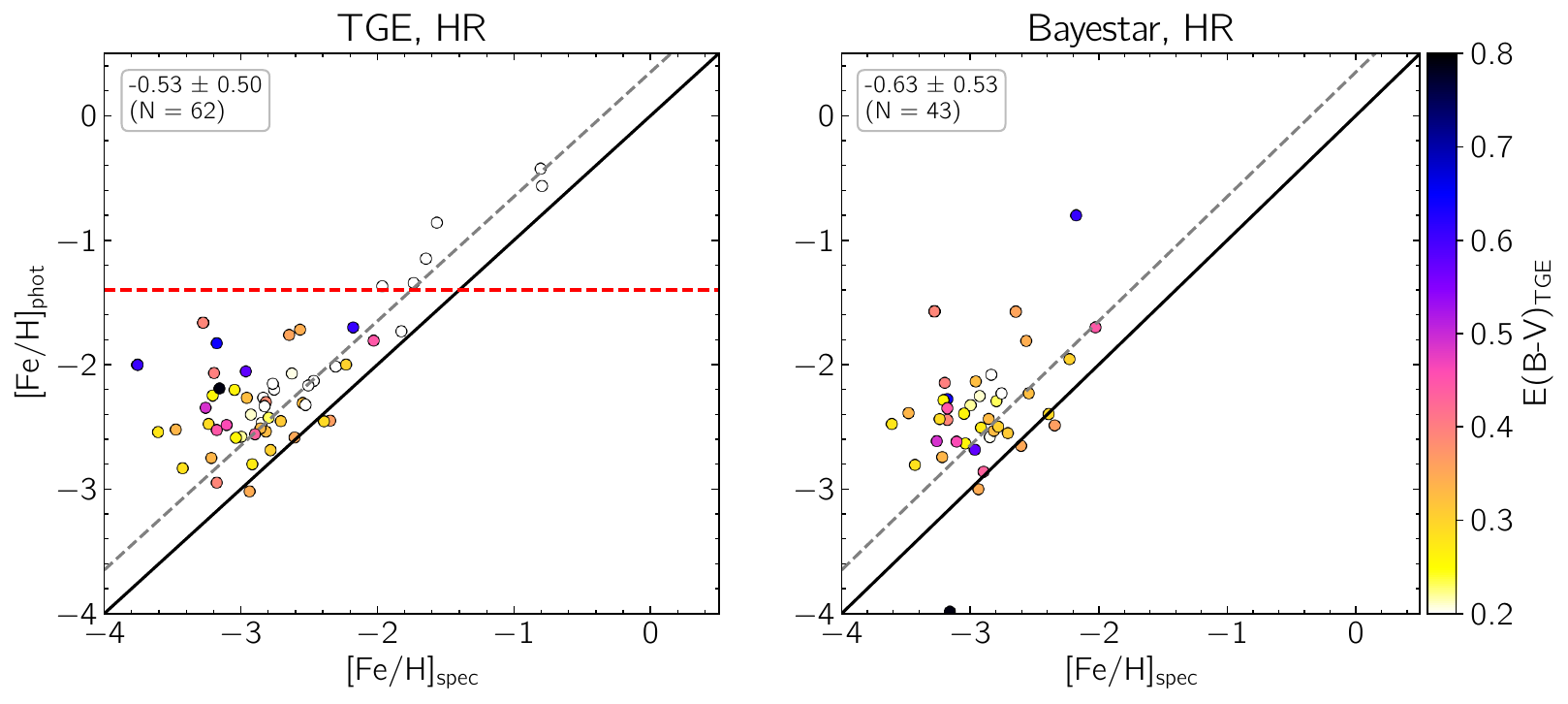}
\caption{Comparison of PIGS photometric vs. spectroscopic metallicities from APOGEE (\textbf{top}), PIGS (\textbf{middle}) and high-resolution literature from \citet{howes14, howes15, howes16, lucey19, lucey22, sestito23} and PIGS team in prep. (\textbf{bottom}). Each is shown for two different assumptions for the reddening: the TGE map (\textbf{left}) and the Bayestar map (\textbf{right}). All photometric quality cuts from Section~\ref{sec:quality} have been applied. VMP APOGEE stars re-analysed by \citet{montelius26} are shown in large symbols, using their [M/H]. CEMP stars in PIGS are shown with star symbols, while CEMP stars are excluded in the HR comparison.
All panels are colour-coded by E(B$-$V)$_\mathrm{TGE}$ to remain consistent, and not all stars have Bayestar reddening available so there are slightly more stars in the left-hand panels. 
Sgr dwarf galaxy stars have been removed, see Figure~\ref{fig:spec_sgr} for those stars. 
The solid black line indicates the 1-1 line, the diagonal dashed grey line is offset by $+0.35$~dex. The TGE panels also have the 4MOST selection photometric metallicity indicated by a horizontal red-dashed line, at $\fehphot = -1.4$. 
Numbers give the 3$\sigma$-clipped median and standard deviation of [Fe/H]$_\mathrm{spec} - $[Fe/H]$_\mathrm{phot}$, excluding [Fe/H]$_\mathrm{phot} > -0.1$.
}
\label{fig:spec} 
\end{figure*}

\begin{figure*}
\centering
\includegraphics[width=0.8\hsize,trim={0.0cm 0.0cm 0.0cm 0.0cm}]{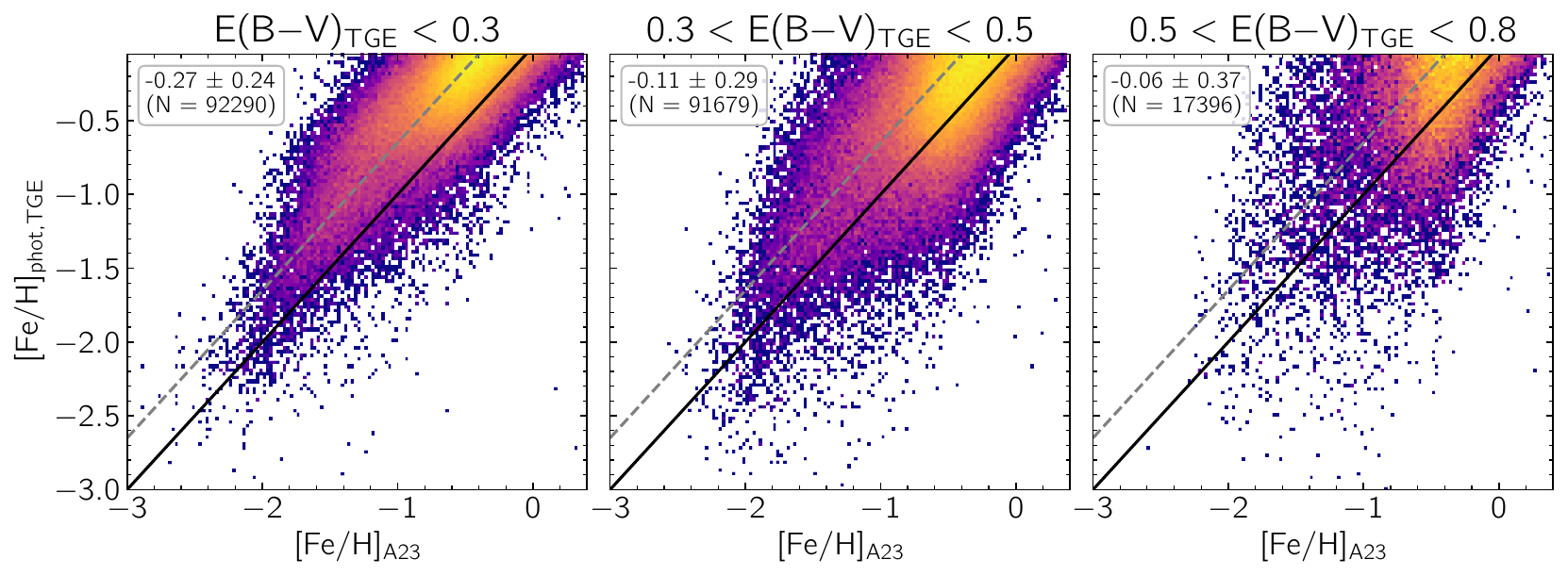}
\caption{Comparison of PIGS TGE photometric metallicities with \Gaia XP XGBoost metallicities from the vetted RGB sample of \citet{andrae23}, in three bins of E(B$-$V). The colour-coding is logarithmic density. The solid black line indicates the 1-1 line, the diagonal dashed grey line is offset by $+0.35$~dex. 
Numbers give the 3$\sigma$-clipped median and standard deviation of [Fe/H]$_\mathrm{A23} - $[Fe/H]$_\mathrm{phot}$, excluding [Fe/H]$_\mathrm{phot} > -0.1$.
}
\label{fig:a23} 
\end{figure*}

\subsubsection{APOGEE \& PIGS}\label{sec:apopigs}
Here we compare our photometric metallicities with spectroscopic metallicities from the high-resolution spectroscopy survey APOGEE DR17 \citep{apogeedr17}, supplemented in the VMP regime with the dedicated analysis by \citet{montelius26}, and the low/medium resolution spectroscopic PIGS follow-up \citep{arentsen20_II, arentsen24}. The APOGEE data covers a wide range in metallicity, making it useful for testing the higher metallicity regime as well as potential contamination. The training sample for the \Pristine photometric metallicity code contained APOGEE stars (shifted down by 0.1~dex to match the scale of the rest of the training sample), but all stars with E(B$-$V)$_\mathrm{SFD}$ $> 0.1$ were removed (\citetalias{martin24}). Stars in the reddened inner Galaxy are therefore still a good test sample. We keep all APOGEE stars with SNR $>50$\footnote{We experimented with additional cuts but no additional stars were excluded.}. The APOGEE sample magnitude distribution peaks around G~$=15$. The spectroscopic PIGS sample is focused on low metallicity stars, although it also has a slight tail towards higher metallicity\footnote{Note that for the previous PIGS spectroscopic follow-up, we did not have photometric metallicities yet. Instead, we used a relative selection in each AAT AAOmega+2dF field, selecting the $\sim 350$ stars highest in the extinction-corrected \Pristine colour-colour diagram, where the most metal-poor stars are expected to lie. Details are described in \citet{arentsen20_II}.}. The PIGS spectra were analysed with two different pipelines \citep{arentsen20_II}, we adopt the \FERRE values as they are recommended for metal-poor stars. We select all stars with the recommended quality cuts, by only keeping stars with \texttt{good\_ferre} = True, and SNR\_4000\_4100 $>20$ (note that this is a stricter SNR cut than usually used in previous PIGS work to ensure high-quality metallicities for the comparison). The magnitude distribution of the resulting spectroscopic PIGS sample peaks around G~$=16$. 

Figure~\ref{fig:spec} presents the metallicity comparisons with APOGEE (top) and PIGS (middle), for the two different extinction maps that we use in this work -- TGE (left) and Bayestar (right). We apply all the photometric metallicity cuts in the previous sub-section, and remove stars in the Sgr dwarf galaxy. The stars are colour-coded by E(B$-$V), clearly highlighting that the scatter increases for stars with higher extinction, as well as indicating that there are some systematics with extinction. Overall, there is a good correlation, although there is an offset. 

We summarise the median and standard deviation for spectroscopic minus photometric metallicity for stars with [Fe/H]$_\mathrm{phot,TGE} < -0.1$ (avoiding the grid edge) in Table~\ref{tab:diff}. Based on these results, for the 4MOST selection (see Section~\ref{sec:4most}) we decided to use the TGE metallicities minus 0.35~dex (referred to as \feh$_\mathrm{phot,TGE,corr}$). 

We also tested stars with $\texttt{merge\_flag} = +2$ and $+1$. For $+2$, there are 120 and 60 stars in common with PIGS and APOGEE, respectively, and only 8 and 32 for $+1$, respectively. The offsets are similar but the standard deviations appear to be slightly higher -- for example, for the APOGEE data in the $0.3 <$ E(B$-$V) $<0.5$ bin, the median and standard deviations for the TGE metallicities are $-0.3 \pm 0.35$ for $+2$ and $-0.26 \pm 0.37$ for $+1$. 

We test the dependence on \teff in Appendix Figure~\ref{fig:spec_teff}. We find a slight trend with \teff, which is strongest for the warmest stars in PIGS with $\teff \gtrsim 5700$~K -- these are horizontal branch stars. Between 4500~K and 5700~K there is a difference of $\sim 0.1$--$0.2$~dex, rising by an additional $\sim0.2$~dex between 5700~K and 6000~K.

Carbon-enhanced metal-poor (CEMP) stars, with [C/Fe] $> +0.7$ from the \FERRE analysis of the PIGS spectra and corrected for evolutionary effects \citep{arentsen21}, are shown with large star symbols in Figure~\ref{fig:spec}. They are clearly offset from the correlation. This is not surprising given the impact of large molecular carbon bands on the spectrum (see also \citetalias{martin24} and \citealt{yuan26}). 

\begin{table}
\centering
\caption{\label{tab:diff} Median and standard deviation for the spectroscopic vs. photometric metallicity comparisons, in different extinction bins. Stars with [Fe/H]$_\mathrm{phot,TGE}>-0.1$ have been removed because they are close to the grid edge.
}
\begin{tabular}{lccc}
 \hline
 \textbf{E(B$-$V)$_\mathrm{TGE}$ =} & $0-0.3$ & $0.3-0.5$ & $0.5-0.8$ \\
 \hline
 \multicolumn{3}{|l|}{\textbf{$\mu \pm \sigma$ (PIGS $-$ phot)} } & \\
 \hline
TGE & $-0.36 \pm 0.21$ & $-0.29 \pm 0.27$ & $-0.46 \pm 0.41$ \\
Bayestar & $-0.37 \pm 0.22$ & $-0.37 \pm 0.25$ & $-0.37 \pm 0.36$ \\
 \hline
 \multicolumn{3}{|l|}{\textbf{$\mu \pm \sigma$ (APOGEE $-$ phot)}} & \\
 \hline
TGE & $-0.34 \pm 0.22$ & $-0.28 \pm 0.25$ & $-0.2 \pm 0.46$ \\
Bayestar & $-0.22 \pm 0.25$ & $-0.24 \pm 0.28$ & $-0.1 \pm 0.48$ \\
\\
\textbf{[Fe/H]$_\mathrm{spec}$ $< -1$:} & & & \\
TGE & $-0.43 \pm 0.24$ & $-0.31 \pm 0.24$ & $-0.46 \pm 0.43$ \\
Bayestar & $-0.32 \pm 0.24$ & $-0.27 \pm 0.28$ & $-0.2 \pm 0.52$ \\
 \hline
 \multicolumn{3}{|l|}{\textbf{$\mu \pm \sigma$ (\citetalias{andrae23} $-$ phot)}} & \\
 \hline
\textbf{mean [Fe/H] $< -1$:} & & & \\
TGE      & $-0.32 \pm 0.31$ & $-0.15 \pm 0.44$ & $-0.01 \pm 0.60$ \\
Bayestar & $-0.26 \pm 0.34$ & $-0.18 \pm 0.44$ & $0.04 \pm 0.56$ \\
 \hline
\end{tabular}
\end{table}

\subsubsection{High-resolution literature}\label{sec:hrspec}

To test the quality of our photometric metallicities in the very metal-poor regime, we compile a sample of literature high-resolution spectroscopic measurements. All literature stars have been pre-selected from narrow-band photometry, either from PIGS (\citealt{sestito23}, UVES follow-up by PIGS team in prep.) or SkyMapper \citep{howes14, howes15, howes16, lucey19, lucey22}. The PIGS-UVES follow-up will be described in a future work, but in short, the \teff was derived by fitting H-alpha with a non-LTE profile interpolated in the \citet{amarsi18} grid, \logg was selected to be consistent with an $\feh = -3.0$ isochrone while having consistent Fe~I and Fe~II abundances in non-LTE, and \feh was derived from Fe-II lines. 

The comparison is presented in the bottom panels of Figure~\ref{fig:spec}. The agreement for low-extinction stars is excellent, with the same $\sim 0.35$~dex offset as identified for PIGS and APOGEE. The scatter increases for stars with higher extinction, which is not unexpected. There is also a potential bias towards higher photometric metallicities at high extinction. In Section~\ref{sec:interstellar}, we consider whether some of this bias could be due to interstellar \CaHK. 

\begin{figure}
\centering
\includegraphics[width=0.9\hsize,trim={0.0cm 0.0cm 0.0cm 0.0cm}]{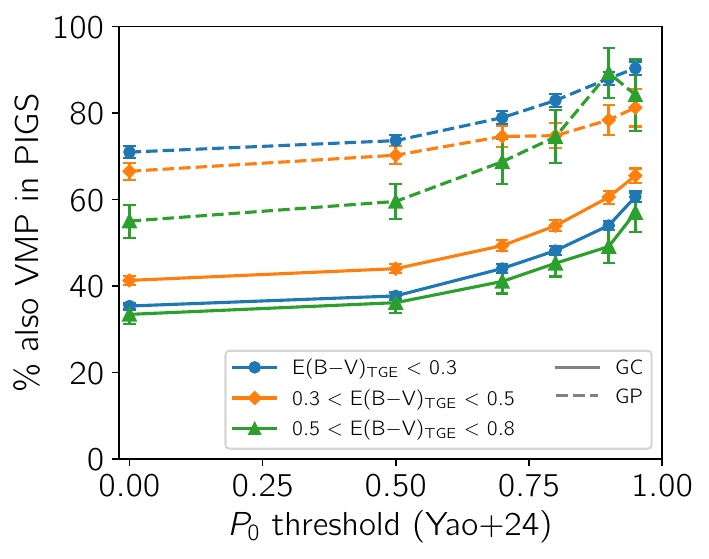}
\caption{Fraction of \citet{yao24} giant VMP candidates that are also classified as VMP by PIGS ([Fe/H]$_\mathrm{phot,TGE,corr}$ $<$ $-2.0$), as a function of the \citet{yao24} VMP probability threshold $P_0$. The result is split in three extinction bins. Solid lines show the high-completeness (GC) classifier and dashed lines the high-purity (GP) classifier. Error bars represent $1\sigma$ binomial uncertainties.}
\label{fig:yao} 
\end{figure}

\begin{figure*}
\centering
\includegraphics[width=0.8\hsize,trim={0.0cm 0.0cm 0.0cm 0.0cm}]{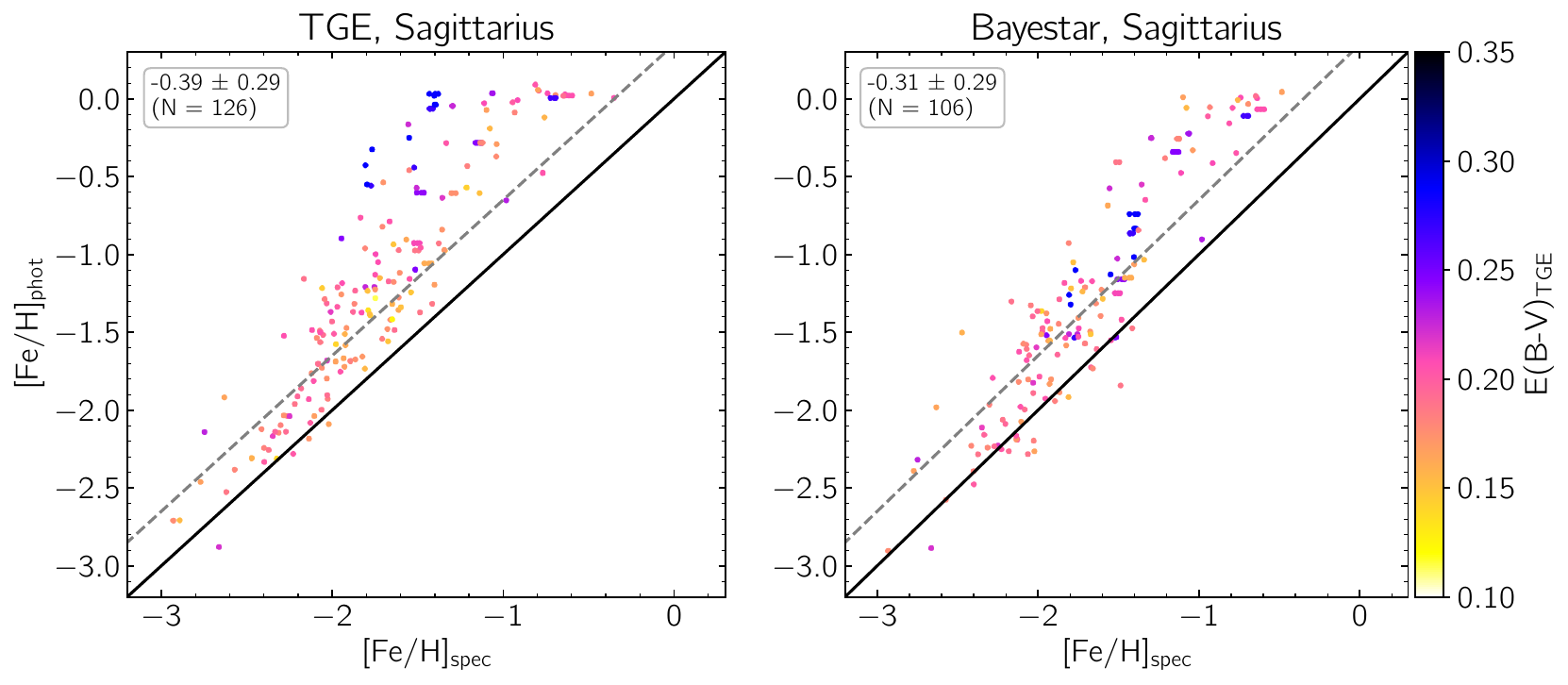}
\caption{Comparison between PIGS photometric metallicities and PIGS and APOGEE spectroscopic metallicities for stars selected to be in the Sgr dwarf galaxy, analogous to Figure~\ref{fig:spec} (although note the different x-, y- and colour-bar limits). Stars have been included with (BP$-$RP)$_0$ up to 1.4 instead of 1.3, as most metal-rich Sgr stars in APOGEE would have been excluded otherwise. 
Numbers give the 3$\sigma$-clipped median and standard deviation of [Fe/H]$_\mathrm{spec} - $[Fe/H]$_\mathrm{phot}$, excluding [Fe/H]$_\mathrm{phot} > -0.1$.
}
\label{fig:spec_sgr} 
\end{figure*}

\subsubsection{Gaia XP metallicities}

\Gaia XP spectra can also be used to derive metallicities, and a much-used sample is that from \citet[][hereafter A23]{andrae23} who trained XGBoost on APOGEE and the \citet{li22} sample of very metal-poor stars to predict metallicities. We compare our TGE metallicities to their predictions for their vetted RGB sample in Figure~\ref{fig:a23}, and summarise the results for both TGE and Bayestar in the bottom rows of Table~\ref{tab:diff}. Unsurprisingly, the results look similar to our APOGEE comparison in the low-extinction regime. However, at higher extinction, a strong bias begins to appear ($+0.1$ to $+0.3$~dex difference with the lowest extinction bin), which is different from that arising in the PIGS comparison with the spectroscopic samples. This suggests the \citetalias{andrae23} metallicities for metal-poor stars in high-extinction regions could be systematically over-estimated. This is supported by Appendix Figure~\ref{fig:a23_lit}, where we compare the \citetalias{andrae23} metallicities with the same literature samples as used in Figure~\ref{fig:spec}. Some of this could also be due to limited sensitivity for fainter stars. Investigating the full nature of these biases is beyond the scope of this work, but potential biases at low metallicity in high extinction regions should be considered when using this or similar datasets in regions like the inner Galaxy. 

Another set of metallicities based on \Gaia XP is the \texttt{StarHorse} \texttt{XGBoost}, or \texttt{SHBoost}, catalogue by \citet{khalatyan24}, which is used for the target selection of the VMP sub-survey in 4MIDABLE-LR. Appendix Figure~\ref{fig:sh_lit} shows the comparison of these metallicities with the spectroscopic reference sets used in this work. As discussed in \citet{khalatyan24} and also shown in \citet{thai26}, SHBoost tends to overestimate the metallicity for metal-poor stars, which can be seen in the figure. The 4MIDABLE-LR VMP survey uses the selection $\texttt{xgbdist\_met\_mean} + \texttt{xgbdist\_met\_std} < -1.0$, which should still result in a clean selection of (very) metal-poor stars.

We also compare our results to the \citet{yao24} catalogue of VMP candidates from \Gaia XP, which provides probabilities $P_0$ that stars have $\feh < -2.0$ for two giant catalogues: high-completeness (GC) and high-purity (GP). Figure~\ref{fig:yao} shows the fraction of \citet{yao24} candidates also classified as VMP by PIGS ($\feh_\mathrm{phot,TGE,corr} < -2.0$) as a function of $P_0$, split by extinction. The GP catalogue shows better agreement with PIGS by a factor of 1.5--2, though GC contains 3--4 times more stars in total. Agreement improves with increasing $P_0$ and decreasing extinction, and is likely somewhat underestimated due to scatter in the PIGS metallicities around the $-2.0$ threshold. Overall, the two catalogues show reasonable agreement, and the \citet{yao24} catalogue can be used to select reliable VMP candidates even at higher extinction, provided a sufficiently high $P_0$ threshold is applied.

\subsubsection{Sagittarius stars}
Both APOGEE and PIGS have done dedicated follow-up of Sgr dwarf galaxy stars, so we can compare to those separately. The following simple selection was made to select Sgr stars: proper motions within $0.45''$ of \texttt{pmra} = $-2.68''$ and \texttt{pmdec} = $-1.38''$ (based on a visible proper motion over-density in the spectroscopic samples), parallax $-$ parallax\_error $< 0.1''$ (selecting distant stars), and RA $>280$~deg (where the main body of Sgr is located). The results are shown in Figure~\ref{fig:spec_sgr}. The Bayestar map appears to perform better in this regime. This could potentially be because the TGE map resolution is lower in this part of the sky compared to closer to the Galactic plane. 
Although there are not many stars in the more metal-rich regime, there appears to be a stronger bias for them compared to Figure~\ref{fig:spec}, e.g. in the Bayestar comparison at $\feh_\mathrm{spec} = -1.0$ the Sgr stars end up at $\fehphot = -0.25$ compared to $-0.65$ for the inner Galaxy stars, which we speculate is because stars in Sgr have much lower alpha abundances in this regime than inner Galaxy or typical halo stars -- see also the discussion in \citet{vitali22}. 

\subsection{The persistent metallicity offset}

The $\sim0.35$~dex offset between our photometric metallicities and spectroscopic metallicities is consistent between APOGEE, PIGS and a collection of high-resolution samples, and is only mildly extinction-dependent. It is unclear where this relatively large offset is coming from. 

When cross-matching the main \Pristine survey photometric metallicities from \citetalias{martin24} with APOGEE and applying the relevant quality cuts, there is a narrow correlation ($\sigma=0.12$~dex) with an offset of $0.05$~dex (in the other direction!). The $\sim0.35$~dex offset is therefore not inherited from the main survey calibration. 

The introduced extinction fudge factor should not be the culprit either, as a fudge factor of 1 (meaning none) in Figure~\ref{fig:fudge} still leads to a large negative offset, even for the lower extinction bins. Additionally, the fact that both the TGE and Bayestar results converge on the same metallicity offset is striking. Potentially something else in the iterative extinction correction in the photometric metallicity code could cause systematics. However, the main \Pristine survey, which uses the same code, does not find significant systematic metallicity offsets up to E(B$-$V)$_\mathrm{SFD}$ $\sim 0.3$ (less than $0.05$~dex, \citetalias{martin24}). The systematics they do find at higher E(B$-$V) are likely due to the limitations of the SFD map, and are still only $\sim 0.1$~dex at E(B$-$V) $= 0.4$.

Contributions from interstellar \CaHK are also unlikely to be the driver of the offset, with typical effects of the order of $\sim 0.05$~dex -- see the next section for details. This effect does bring down the photometric metallicities, hence bringing them slightly closer to the spectroscopic ones. 

A difference in the calibration of the \CaHK photometry between the inner Galaxy and the Galactic halo due to stars having different magnitudes and colours is likely also not the cause, as explored in Appendix Figure~\ref{fig:cal2}. Potentially there are other spatial systematics independent of this in the \Gaia XP spectra that are stronger in the inner Galaxy, which could affect the synthetic \CaHK derivation and therefore the PIGS calibration. However, they would have to be of the order of $\sim 0.1$~mag in \CaHK to result in a 0.35~dex metallicity offset\footnote{This order of magnitude can be estimated by eye from Figure~\ref{fig:ccd}, but we also reran the photometric metallicity code for the PIGS spectroscopy sample with different fixed offsets $\Delta$\CaHK, which results in a zero offset for $\Delta$\CaHK = 0.09. Note that for this test we varied $\Delta$\CaHK and \texttt{bay\_fudge} together, and for $\Delta$\CaHK = 0.09 the best \texttt{bay\_fudge} is still 1.1.}, which is too large. 

With no other solutions at the present, we decide to correct for this metallicity offset when selecting or counting metal-poor giant stars from PIGS photometry in the regime we are interested in here. However, it should not necessarily be taken as a universal correction factor for all PIGS photometric metallicities.  

\section{The effect of interstellar Ca H\&K on photometric metallicities}
\label{sec:interstellar}

On its journey towards us, light from inner Galaxy stars encounters the interstellar medium (ISM). Several ISM absorption features can be seen in stellar and extragalactic spectra, such as diffuse interstellar bands (previously explored in the PIGS spectroscopic follow-up in \citealt{zhao23}) and narrow lines from elements like sodium and calcium. In fact, the two main lines relied on in this work to determine photometric metallicities, Ca H\&K, also have an interstellar component. These can be resolved with high-resolution spectroscopy, and more easily for very metal-poor stars with weak stellar lines and velocities different from the ISM. At high Galactic latitudes, they have also been probed using low-resolution SDSS extragalactic spectra \citep{murga15}, which have no or weak `intrinsic' Ca H\&K features. 

There have not previously been any tests of the effect of interstellar Ca H\&K on photometric metallicities (although \citealt{murga15} do note that it should be taken into account for precision photometry). This is predominantly because compared to the stellar lines, the ISM component is very small towards the Galactic halo, where most of our metal-poor surveys have focused. But what about the high extinction inner Galaxy? Could the PIGS photometric metallicities, especially towards the extremely metal-poor tail, be biased due to large interstellar Ca H\&K contributions? 

\subsection{Data and models}

In this work, we address this question by determining the strength of interstellar Ca H\&K lines from high-resolution VLT/UVES spectra of extremely metal-poor PIGS stars. The data will be described in detail in a future work, but very briefly: we observed 22 very metal-poor candidates selected from the PIGS/AAT analysis, with the goal of determining detailed chemical abundances from the R$\sim 41\,000$ spectra. The S/N around Ca H\&K is in the range of 15-30, sufficient to test the contribution of interstellar Ca H\&K for all stars (apart from one star with S/N=10). They have spectroscopic metallicities in the range $-3.5 < \feh < -2.5$ and are all red giant branch stars. Using \texttt{IRAF} \citep{iraf}, we interactively normalise the spectra in the $3800-4100$~\AA range with a cubic spline of varying order, depending on the spectrum. 

We use the adopted stellar parameters for the sample (briefly described in Section~\ref{sec:hrspec}) to create small grids of synthetic spectra for each UVES star, with offsets [$-150, -75, 0, +75, +150$] K around the adopted \teff and [$-0.3, -0.2, -0.1, 0.0, +0.1, +0.2, +0.3$] around the adopted \feh. We create them from the \texttt{hnsc1} grid by \citet{allende18} using the interpolation function in \FERRE\footnote{\FERRE is available from \url{http://github.com/callendeprieto/ferre}} \citep{allende06}. This grid assumes [$\alpha$/Fe] = $+0.4$ for very metal-poor stars, which should be appropriate for this sample, but any variations can also be absorbed by the use of our small grid in \feh. We cut the synthetic spectra to $3800-4100$~\AA and smooth them to $R=41\,000$. 

We exclude six stars with $\teff \leq 4600$~K, a regime where the model spectra are not good enough for our present purpose, as well as one star strongly enhanced in carbon \citep{mashonkina23}. After these cuts, there remain two stars for which we are unable to measure interstellar CaHK: the star with the lowest S/N, and a highly reddened star that does not show narrow interstellar CaHK features. The synthetic model for this star does not match the stellar CaHK features well, which may indicate broad interstellar absorption features that we are not able to characterise with our adopted method. The high reddening of this star, E(B$-$V) = 0.7, may correspond to more complex interstellar CaHK features.  

\subsection{Analysis}

\begin{figure}
\centering
\includegraphics[width=0.9\hsize,trim={0.0cm 0.0cm 0.0cm 0.0cm}]{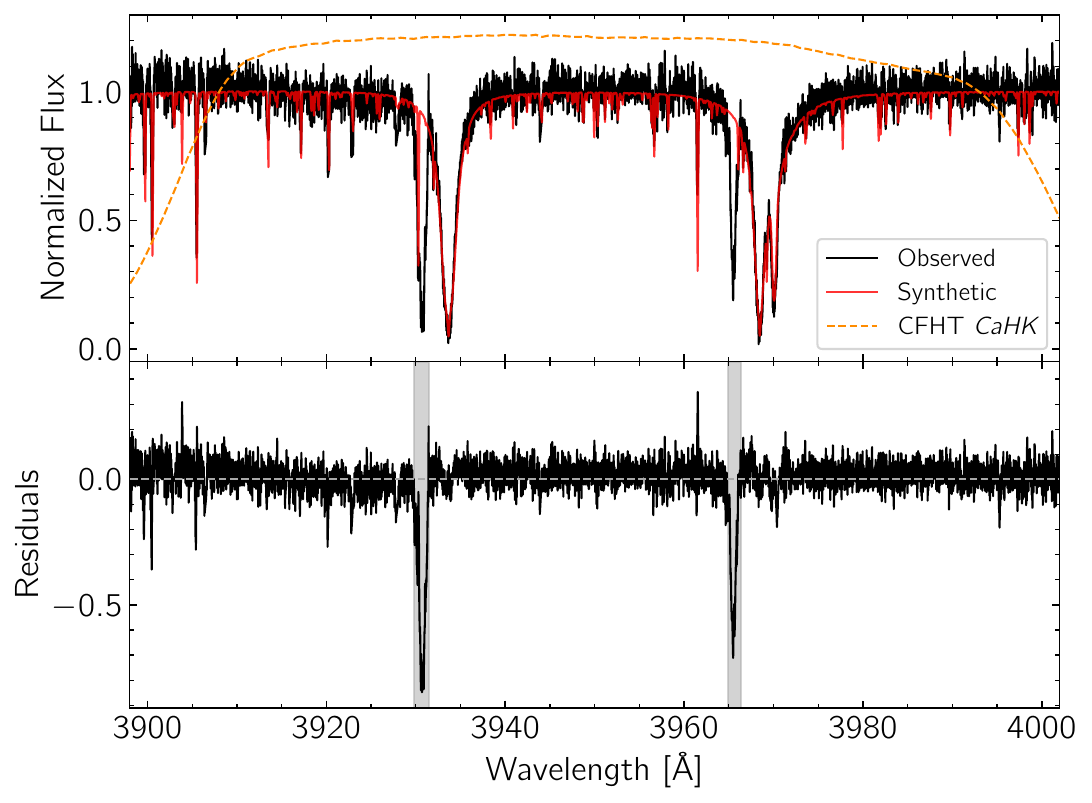}
\includegraphics[width=0.9\hsize,trim={0.0cm 0.0cm 0.0cm 0.0cm}]{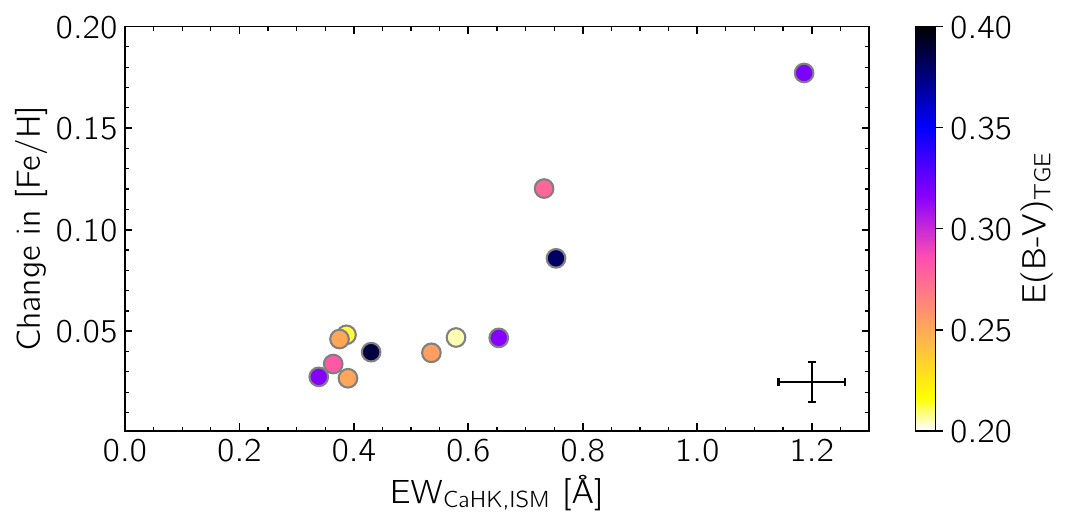}
\caption{Top panel, first row: observed UVES spectrum (black) and its best synthetic spectrum (red) for the star with the highest Ca H\&K ISM EW. The CFHT \CaHK filter curve is shown in the orange-dashed line. Top panel, second row: residual, showing the interstellar lines only. The grey-shaded regions indicate the regions used to compute the ISM EW. Bottom panel: change in photometric metallicity as function of the ISM EW, colour-coded by E(B$-$V). Typical uncertainties are indicated in the bottom right corner. 
}
\label{fig:ism} 
\end{figure}

To isolate the Ca H\&K ISM contribution for each star, we take the difference between the observed and best synthetic spectrum in our small grid for that star, which is identified using the minimum $\chi^2$. An example is shown in the top panel of Figure~\ref{fig:ism}. We measure the equivalent widths for the ISM Ca K and Ca H features separately in each residual spectrum using the \texttt{equivalent\_width} function in the \texttt{specutils}\footnote{\url{https://zenodo.org/records/14042033}} Python package. As a sanity check we also measure them by hand using IRAF. The sum of Ca H\&K agrees well, with a median of $-0.030$ and a standard deviation of $0.082$~\AA. We use the standard deviation divided by $\sqrt{2}$, giving $0.058$~\AA, as an estimate of our EW uncertainties. This does not take into account systematics due to e.g. the continuum placement and the quality of the synthetic spectra, but still provides a rough estimate. 

To assess the effect of these interstellar lines on \Pristine photometric metallicities, we derive a corrected CaHK magnitude CaHK$_\mathrm{cor}$ as follows:

\begin{equation}
\text{CaHK}_\mathrm{cor} = \text{CaHK} - 2.5 \log{\frac{\Delta\lambda\  - \mathrm{EW}_{\text{synth}}}{\Delta\lambda\  - (\mathrm{EW}_{\text{synth}} + \mathrm{EW}_{\text{CaHK,ISM}})} }
\end{equation}

where EW$_\mathrm{CaHK,ISM}$ is the sum of the equivalent widths of the two interstellar lines and EW$_\mathrm{synth}$ is the effective EW of all lines in the best synthetic spectrum between $3910-3990$~\AA ($\Delta\lambda = 80$~\AA), where the CFHT \CaHK filter has most of its transmission -- see the orange-dashed line in the top panel of Figure~\ref{fig:ism}. This approximates the flux difference through the \CaHK filter due to interstellar CaHK lines compared to just stellar absorption lines. We then rerun the photometric metallicity estimation (Section~\ref{sec:photmet}) with this CaHK$_\mathrm{cor}$ value. We estimate uncertainties on the photometric metallicities by generating 100 random draws of the EW assuming a Gaussian error with $\sigma = 0.058$~\AA, and taking the standard deviation of the results. These are of the order of $0.004-0.009$, and we adopt 0.01 as the typical uncertainty to be conservative. 

\subsection{Results}

The bottom panel of Figure~\ref{fig:ism} shows the correlation between the ISM equivalent widths and the change in photometric metallicity, colour-coded by extinction. For most of the stars in this sample, the over-prediction in metallicity due to interstellar CaHK is $\sim 0.05$~dex. For three stars the effect is slightly larger ($0.09$--$0.12$), and it rises to $0.18$~dex for the star with the highest EW$_\mathrm{CaHK,ISM}$. The latter star is the one with the highest temperature (5400~K) in the sample, which exacerbates the impact of an ISM contribution in two ways: it has the weakest stellar Ca H\&K lines and therefore the relative effect of adding interstellar lines is stronger, and the same change in the \CaHK magnitude has a larger effect for warmer stars due to a decreased metallicity sensitivity in that region of the \Pristine colour-colour diagram. We systematically test the dependence on metallicity and temperature using a synthetic grid of spectra in Appendix Figure~\ref{fig:ismgrid}.

We note that there is little correlation between the strength of the interstellar Ca H\&K lines and the extinction in our sample (Spearman $\rho = 0.27$, $p = 0.39$). Although this is based on a small sample, it is consistent with previous results by \citet{murga15}, who found a linear relation between E(B$-$V) \citep[from][]{schlegel98} and EW$_\mathrm{Ca K}$ up until E(B$-$V) $\sim$0.1, which became saturated for 0.1 $<$ E(B$-$V) $<$ 0.2, with some scatter around EW$_\mathrm{Ca K} \sim 0.3$~\AA. 
Potentially there might be a better correlation at high extinction if precise distances and detailed 3D dustmaps were used, neither of which are presently available for stars in the inner Galaxy. 

In summary, with the PIGS/UVES sample we probe 0.2 $<$ E(B$-$V) $<$ 0.4, and find that \Pristine photometric metallicities can be overestimated due to interstellar Ca H\&K to the level of $0.05-0.2$~dex. It is not straightforward to make a correction for this as it does not clearly correlate with E(B$-$V). Larger studies of this kind are necessary to identify other potential correlations and the true extent of the induced scatter, as well as to extend it to higher extinction regions. 

\section{The PIGS very metal-poor survey in 4MOST}\label{sec:4most}

\subsection{Motivation}

Spectroscopy of metal-poor inner Galaxy stars is essential for the study of the early Milky Way, revealing the compositions of stars that trace the progression of early star formation and chemical enrichment. Current work with samples of very metal-poor stars in the inner Galaxy is either limited by small-number statistics at high resolution with detailed abundances \citep[e.g.][]{howes14, howes15, howes16, caseyschlaufman15, koch16, reggiani20, lucey19, lucey22, sestito23} or a relatively low spectroscopic resolution with only metallicities \citep{arentsen20_I, arentsen24} and carbon abundances \citep{arentsen21}. 

To efficiently build up a large sample of spectroscopic observations for metal-poor inner Galaxy stars, we have created the \texttt{VMP\_PIGS} sub-survey within the 4MIDABLE-LR survey \citep{chiappini19} in 4MOST. In it, we will target metal-poor candidates from the PIGS photometry. It will make use of the 4MOST low-resolution spectrograph, with $R\sim 4000 - 7700$ and a broad wavelength coverage of $3700 - 9500$~\AA \citep{4most}. We expect to obtain radial velocities, stellar parameters and chemical abundances from multiple nucleosynthetic channels for $\sim20,000$ metal-poor inner Galaxy stars with $\feh < -1.5$ and $\sim7500$ with $\feh < -2.0$ over the course of the 5-year 4MOST survey, tripling to quadrupling the number of stars in the low-resolution PIGS follow-up \citep{arentsen24}. 
This large sample will be complementary to smaller 4MOST samples at high resolution \citep[4MIDABLE-HR,][]{bensby19}, and to the wider but shallower \texttt{VMP} sub-survey in 4MIDABLE-LR selected from \citet{khalatyan24}.
A great advantage is that the stellar parameters and abundances for \texttt{VMP\_PIGS} will be on the same scale as the rest of 4MOST, including the halo and dwarf galaxies, allowing for homogeneous comparisons across Galactic environments. 

The main other spectroscopic surveys targeting the inner Galaxy now and in the next few years are SDSS-V with the APOGEE spectrograph \citep{sdssv} and MOONS \citep{moons, moonsgal}. Both of these are in the (infra)red, which is an advantage for piercing through the dust and both of these surveys can and will observe closer to the Galactic plane, but a disadvantage for studying the chemistry of very low metallicity stars as various elemental lines start disappearing for $\feh < -2.0$, with only some Mg and Si lines and no more Fe lines available in APOGEE for $\feh < -2.5$ \citep{montelius26}. \Gaia RVS spectroscopy can also be used in the inner Galaxy, for example with improved stellar parameters from convolutional neural networks \citep{guiglion24, nepal26}, although abundance information for very metal-poor stars will remain limited. 

\subsection{Observing strategy}

We aim at a minimum SNR per angstr\"om of 40 in the blue ($4500-4700$~\AA) and 50 around the MgB lines ($5140-5200$~\AA), the same requirements as for the 4MIDABLE-LR chemistry survey. It is expected that for metal-poor giants with this SNR, it should in principle be possible to derive abundances for C, N, Fe, several individual alpha elements (Mg, Ca, Ti, Si), iron-peak (Ni, Cr, Mn) elements, odd-Z elements (Sc, Na) and s-process elements (at least Ba, likely also Y, Sr) from the 4MOST low-resolution spectra, according to the \citet{kordopatis23} tool. There will most likely be at least one element for each of the main nucleosynthetic groups (apart from the r-process). The radial velocity precision is expected to be $\sim1$~\kms, good for high-quality orbit derivations when combined with \Gaia proper motions and spectro-photometric distances. The latter will be derived for all targets in 4MIDABLE-LR with the \texttt{StarHorse} code \citep{Santiago16, Queiroz18}. 

The desired SNR is achievable with 1h (or shorter) exposures for the bulk of our targets based on the 4MOST exposure time calculator (ETC), with stars in the range $14 < G< 17$. We follow the 4MIDABLE-LR strategy in providing synthetic spectroscopic templates for the ETC, which needs a rough estimate of the reddening and stellar parameters. They are dereddened to the nearest 0.2~mag in E(B$-$V) using \texttt{StarHorse} EDR3 estimates \citep{Anders22} if available, to be consistent with the rest of 4MIDABLE-LR, and if not, we use TGE rescaled to the \texttt{StarHorse} values. The template \feh is based on the PIGS photometric metallicity in steps of 0.5~dex, $\alphafe = +0.4$ (appropriate for most metal-poor stars) and template \teff and \logg are based on the \texttt{StarHorse} run where available, in steps of 500~K and 0.5~dex, respectively. For the 7\% of stars where they were not available because \texttt{StarHorse} did not converge, \teff is based on the \Gaia infrared flux method following \citet{casagrande21}  and \logg is fixed to an appropriate value for the given \teff\,(only a rough estimate of \logg is needed, we fix \teff--\logg pairs to overlap with the PIGS/AAT parameter distribution).

We select all targets in the PIGS footprint with $|l|$ and $|b| < 12^\circ$ that pass our quality criteria described in Section~\ref{sec:quality} and have $\feh_\mathrm{phot,TGE,corr} < -1.75$). In the APOGEE and PIGS data, this photometric metallicity cut results in a sample where $\sim90\%$ of the stars have spectroscopic $\feh < -1.5$ and $\sim 20-40\%$ (APOGEE and PIGS, respectively) have $\feh < -2.0$  -- see the left-hand panel of Figure~\ref{fig:spec} under the horizontal dashed line. We note that our selection is biased against CEMP stars\footnote{Our PIGS/AAT follow-up was likely also biased against at least some CEMP stars, see \citet{arentsen21}, but less so than the current selection. This is because there were more fibres available per field than the number of most metal-poor candidates, which were filled with the next best candidates.}. Our final selection results in $30\,413$ metal-poor star candidates, out of 1.2 million before the metallicity cut. The total number of survey targets is higher than the survey goal, to give the 4MOST simulator \citep{tempel20b,tempel20a} flexibility in the target selection -- we consider our survey successful if 2/3rd of the stars ($\sim 20\,000$) are observed fully according to our SNR criteria. 

\begin{figure}
\centering
\includegraphics[width=1.0\hsize,trim={0.0cm 1.0cm 0.0cm 0.0cm}]{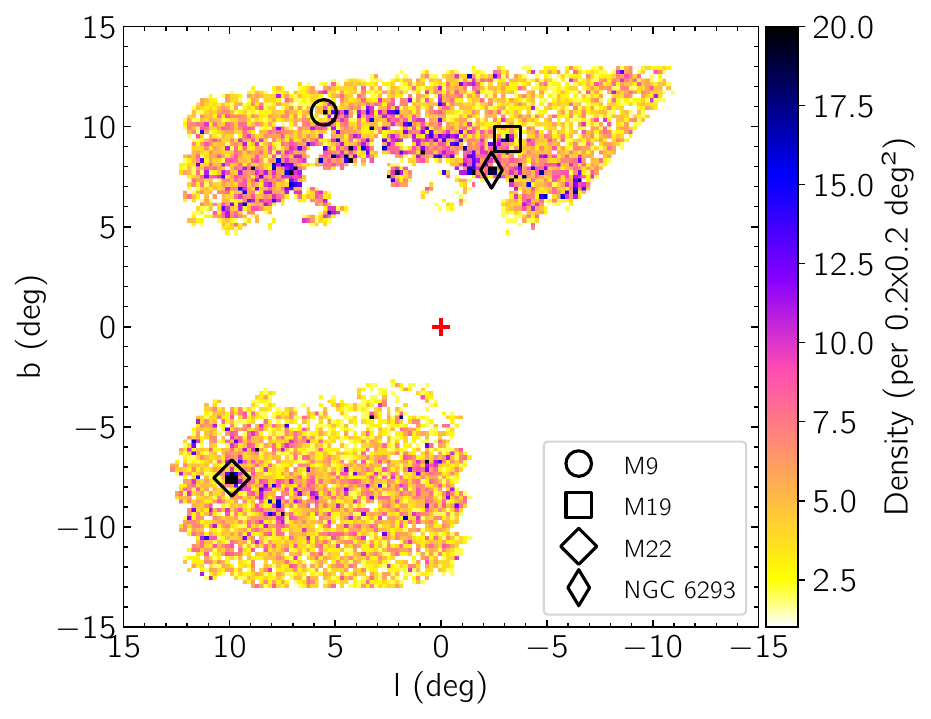}
\caption{Footprint of the 4MIDABLE-LR \texttt{VMP\_PIGS} catalogue of metal-poor stars. The plus symbol denotes the Galactic centre, and the four globular clusters mentioned in the text are highlighted as indicated in the legend.
}
\label{fig:4foot} 
\end{figure}

The distribution of the sample is shown in Figure~\ref{fig:4foot}, with a typical density of the order of $50-200$ per deg$^2$. There are four clear over-densities in the distribution, corresponding to the metal-poor globular clusters M9, M19, M22 and NGC~6293, with reported metallicities in the \citet{harris10} catalogue of $\feh = -1.77, -1.74, -1.7$ and $-1.99$, respectively. They were not specifically selected but we do not remove them from our metal-poor catalogue as it will be of interest to observe them homogeneously with their surroundings. 

The 4MOST simulator tries to build the optimal 4MOST 5-year survey, observing targets from all surveys and sub-surveys simultaneously. Targets do not have to be finished (meaning the required SNR is reached) in one pointing, but could be split over multiple visits. This can be an advantage for probing stellar radial velocity variations, if the observations are separated enough in time. We attempt to go beyond the ``cadence for free'' by requesting that if the simulator decides a target should be split over multiple observations, that these observations are separated in time by at least 300 days. This increases the number of useful radial velocity epochs, but will only be implemented on a best effort basis and is not a key driver for the survey. 

The survey strategy described here will result in a high-quality chemical dataset for ancient inner Galaxy stars, which, combined with \Gaia astrometry, will lead to a valuable chemo-dynamical dataset for dissecting the early Milky Way. 

\section{Summary}\label{sec:summary}

In this work, we present the photometric Pristine Inner Galaxy Survey (PIGS), which covers 340 deg$^2$ towards the Galactic bulge (see Figure~\ref{fig:footprint}). A summary of our results: 

\begin{itemize}
    \item We described the observations, data reduction and calibration (largely following the main \Pristine survey strategy, see \citealt{martin24}) and final catalogue of \CaHK photometry for 55 million stars. The latter is publicly released with this paper, with typical uncertainties of $\lesssim 0.05$ for \CaHK $\sim 21$. 
    \item In order to generate reliable photometric metallicities, it was necessary to apply an empirical correction to the extinction values in our photometric metallicity framework. We used the difference between spectroscopic and photometric metallicities to do this and found convergence for a best ``fudge factor'' for the TGE and Bayestar maps that minimises photometric metallicity trends with E(B$-$V). See Figure~\ref{fig:fudge}. This methodology could in principle be used to study spatial variations of dust properties, but requires more development. 
    \item We release photometric metallicities for all 55 million PIGS stars. These have been validated for $G<17$, caution is required for fainter stars. There are 8.4 million stars with $G < 17$, of which 1.2 million pass our main quality cuts. The distribution of those metallicities is shown in Figure~\ref{fig:mdf}. Stricter cuts are necessary to select stars at the extremely metal-poor end.
    \item We validated the photometric metallicities using spectroscopic data from APOGEE, PIGS and small optical high-resolution samples. There is a good agreement, albeit with a consistent $\sim 0.35$~dex offset, and an increased scatter for stars with higher E(B$-$V). See Figure~\ref{fig:spec}. 
    \item As is the case for the main \Pristine survey, PIGS is over-predicting the metallicity of CEMP stars. This results in a bias against such stars in PIGS-selected very metal-poor samples. 
    \item We compare the \Gaia XP XGBoost metallicities from \citet{andrae23} with our photometric metallicities and with literature spectroscopy, and find that the \Gaia XP XGBoost metallicities appear to be systematically over-predicted for metal-poor stars in high extinction regions. 
    \item We investigated the effect of interstellar Ca H\&K on our photometric metallicities using high-resolution spectroscopy, and find potential biases of the order of $0.05-0.2$~dex, without a clear correlation with E(B$-$V) in our small sample. See Figure~\ref{fig:ism}. 
    \item We presented the survey strategy for the 4MOST follow-up of PIGS metal-poor candidates, the 4MIDABLE-LR/\texttt{VMP\_PIGS} sub-survey, which will result in low-resolution spectra for at least $20,000$ metal-poor inner Galaxy stars. This new dataset will be valuable for chemo-dynamical studies of the ancient Milky Way. 
\end{itemize}

Beyond the spectroscopic follow-up, further work can be done with the photometric PIGS catalogues themselves. Combined with improved extinction corrections and better proper motions and parallaxes from \Gaia DR4, it is a powerful dataset for Galactic Archaeology. 

\section*{Acknowledgements}

We thank the anonymous referee for their helpful comments. We thank Lyudmila Mashonkina and Tatyana Sitnova for their analysis of the PIGS/UVES spectra. 

AAA acknowledges support from the Institute of Astronomy's CamCEAD buyout scheme, the Herchel Smith Fellowship at the University of Cambridge, a Fitzwilliam College research fellowship supported by the Isaac Newton Trust and the Royal Society University Research Fellowship [URF$\backslash$R1$\backslash$261504]. 
This research was supported by the International Space Science Institute (ISSI) in Bern, through ISSI International Team project 540 (The Early Milky Way). 
This work was supported by the Programme National Astro of CNRS/INSU with INP and IN2P3, co-funded by CEA and CNES. 
ES acknowledges funding through VIDI grant "Pushing Galactic Archaeology to its limits" (with project number VI.Vidi.193.093) which is funded by the Dutch Research Council (NWO). This research has been partially funded from a Spinoza award by NWO (SPI 78-411). 
ES also acknowledges funding from the European Research Council (ERC, EARLYMW, Grant number 101170507). Views and opinions expressed are those of the author(s) only and do not necessarily reflect those of the European Union or the European Research Council. Neither the European Union nor the granting authority can be held responsible for them. 
SV acknowledges financial support from the Spanish Ministry of Science, Innovation and Universities (MICIU) project PID2023-149982NB-I00. I.H.A., and from “La Caixa” Foundation (ID 100010434). AV gratefully acknowledges support from the Canadian Institute for Theoretical Astrophysics (CITA) through a CITA National Fellowship and the International Astronomical Union (IAU) and the Gruber Foundation (TGF) through a IAU TGF Fellowship. 

Based on observations obtained with MegaPrime/MegaCam, a joint project of CFHT and CEA/DAPNIA, at the Canada-France-Hawaii Telescope (CFHT) which is operated by the National Research Council (NRC) of Canada, the Institut National des Science de l'Univers of the Centre National de la Recherche Scientifique (CNRS) of France, and the University of Hawaii.

Based on observations collected at the European Southern Observatory under ESO programme 105.20P8.001. 

We thank the Australian Astronomical Observatory, which have made the PIGS spectroscopic follow-up observations used in this work possible. We acknowledge the traditional owners of the land on which the AAT stands, the Gamilaraay people, and pay our respects to elders past and present. Based on data obtained at Siding Spring Observatory (via programs S/2017B/01, A/2018A/01, OPTICON 2018B/029 and OPTICON 2019A/045, PI: A. Arentsen and A/2020A/11, PI: D. B. Zucker). 

This work has made use of data from the European Space Agency (ESA) mission {\it Gaia} (\url{https://www.cosmos.esa.int/gaia}), processed by the {\it Gaia} Data Processing and Analysis Consortium (DPAC, \url{https://www.cosmos.esa.int/web/gaia/dpac/consortium}). Funding for the DPAC has been provided by national institutions, in particular the institutions participating in the {\it Gaia} Multilateral Agreement.

Funding for the Sloan Digital Sky Survey IV has been provided by the Alfred P. Sloan Foundation, the U.S. Department of Energy Office of Science, and the Participating Institutions. SDSS-IV acknowledges support and resources from the Center for High Performance Computing  at the University of Utah. The SDSS website is www.sdss4.org. SDSS-IV is managed by the 
Astrophysical Research Consortium 
for the Participating Institutions 
of the SDSS Collaboration including 
the Brazilian Participation Group, 
the Carnegie Institution for Science, 
Carnegie Mellon University, Center for 
Astrophysics | Harvard \& 
Smithsonian, the Chilean Participation 
Group, the French Participation Group, 
Instituto de Astrof\'isica de 
Canarias, The Johns Hopkins 
University, Kavli Institute for the 
Physics and Mathematics of the 
Universe (IPMU) / University of 
Tokyo, the Korean Participation Group, 
Lawrence Berkeley National Laboratory, 
Leibniz Institut f\"ur Astrophysik 
Potsdam (AIP),  Max-Planck-Institut 
f\"ur Astronomie (MPIA Heidelberg), 
Max-Planck-Institut f\"ur 
Astrophysik (MPA Garching), 
Max-Planck-Institut f\"ur 
Extraterrestrische Physik (MPE), 
National Astronomical Observatories of 
China, New Mexico State University, 
New York University, University of 
Notre Dame, Observat\'ario 
Nacional / MCTI, The Ohio State 
University, Pennsylvania State 
University, Shanghai 
Astronomical Observatory, United 
Kingdom Participation Group, 
Universidad Nacional Aut\'onoma 
de M\'exico, University of Arizona, 
University of Colorado Boulder, 
University of Oxford, University of 
Portsmouth, University of Utah, 
University of Virginia, University 
of Washington, University of 
Wisconsin, Vanderbilt University, 
and Yale University.

This research made extensive use of the \textsc{matplotlib} \citep{matplotlib}, \textsc{pandas} \citep{pandas}, \textsc{astropy} \citep{astropy13, astropy18} and \textsc{dustmaps} \citep{dustmaps} Python packages, and of \textsc{Topcat} \citep{topcat}. 

\section*{Data Availability}

The catalogue described in Table~\ref{tab:photmet_columns} is published alongside this paper via Zenodo: \url{https://doi.org/10.5281/zenodo.24562169}. The PIGS/UVES spectroscopic analysis results will be released in an upcoming paper. All other data  used is publicly available. 

\section*{Author Contributions Statement}

AAA is the PI of PIGS and led all aspects of this work apart from specific ones mentioned next. ES and NFM are co-PIs of the \Pristine survey and contributed to the design and execution of PIGS. NFM handled the data reduction. ZY performed the photometric calibration. AI developed the code for the ISM analysis. CC and IM are co-PIs of 4MIDABLE-LR and contributed to the \texttt{VMP\_PIGS} survey strategy. All other authors are active members of the \Pristine collaboration and took part in discussions and the writing and editing of the paper. ChatGPT was used to draft the abstract (which the authors edited), all other writing is that of the authors. ChatGPT, GitHub Copilot and Claude assisted in table handling and plotting in Python and with the analysis for Figure~\ref{fig:ismgrid}, and Claude performed the final code checks regarding extinction and helped with their interpretation. 

\bibliographystyle{mnras}
\bibliography{ref} %

\appendix

\section{Data release description}

\begin{table*}
\caption{Description of the PIGS data release, for all 55 million PIGS sources with a cross-match in \Gaia DR3 and \CaHK uncertainties $<0.3$.}
\label{tab:photmet_columns}
\centering
\begin{tabular}{l l c c}
\hline\hline
Column & Description & Unit & Type \\
\hline
\texttt{source\_id} & \Gaia DR3 \texttt{source\_id} & & longint \\
\texttt{RA} & \Gaia DR3 right ascension & degrees & float \\
\texttt{Dec} & \Gaia DR3 declination & degrees & float \\
\texttt{RA\_Pr} & PIGS right ascension  & degrees & float\\
\texttt{Dec\_Pr} & PIGS declination  & degrees & float\\
\texttt{angDist} & PIGS--\Gaia DR3 cross-match distance & arcsec & float \\
\texttt{CaHK\_Pr} & PIGS \CaHK magnitude & mag & float \\
\texttt{d\_CaHK} & uncertainty on the PIGS \CaHK magnitude & mag & float \\
\texttt{merged\_CASU\_flag} & Morphology and quality flag & & int \\
\hline
\texttt{ebv\_TGE} & $A_\mathrm{0,TGE} \times \mathrm{\texttt{tge\_fudge}} \times (A_V/A_0) / 3.1 $ & mag & float \\
\texttt{G\_0\_TGE} & De-reddened Gaia $G$ magnitude using TGE & mag & float \\
\texttt{BP\_0\_TGE} & De-reddened Gaia $G_{\rm BP}$ magnitude using TGE  & mag & float \\
\texttt{RP\_0\_TGE} & De-reddened Gaia $G_{\rm RP}$ magnitude using TGE  & mag & float \\
\texttt{CaHK\_0\_TGE} & De-reddened Pristine CaHK magnitude using TGE  & mag & float \\
\texttt{FeH\_Pristine\_TGE} & Photometric metallicity calculated directly from \texttt{CaHK\_0\_TGE} & dex & float \\
\texttt{FeH\_Pristine\_TGE\_16th} & Lower bound (16th percentile) of the TGE photometric metallicity PDF & dex & float \\
\texttt{FeH\_Pristine\_TGE\_50th} & Median value (50th percentile) of the TGE photometric metallicity PDF  & dex & float \\
\texttt{FeH\_Pristine\_TGE\_84th} & Upper bound (84th percentile) of the TGE photometric metallicity PDF  & dex & float \\
\texttt{mcfrac\_Pristine\_TGE} & Fraction of MC samples inside model colour-colour space using TGE & & float \\
\texttt{flag\_good\_TGE} & 1 if all quality cuts described in Section~\ref{sec:quality} are met for the TGE photometric metallicity & & int \\
\hline 
\texttt{ebv\_bay} & E(B$-$V)$_\mathrm{Bayestar} \times \mathrm{\texttt{bay\_fudge}} \times 0.86$ & mag & float \\
\texttt{G\_0\_bay} & De-reddened Gaia $G$ magnitude using Bayestar & mag & float \\
\texttt{BP\_0\_bay} & De-reddened Gaia $G_{\rm BP}$ magnitude using Bayestar & mag & float \\
\texttt{RP\_0\_bay} & De-reddened Gaia $G_{\rm RP}$ magnitude using Bayestar & mag & float \\
\texttt{CaHK\_0\_bay} & De-reddened Pristine CaHK magnitude using Bayestar & mag & float \\
\texttt{FeH\_Pristine\_bay} & Photometric metallicity calculated directly from \texttt{CaHK\_0\_bay}  & dex & float \\
\texttt{FeH\_Pristine\_bay\_16th} & Lower bound (16th percentile) of the Bayestar photometric metallicity PDF & dex & float \\
\texttt{FeH\_Pristine\_bay\_50th} & Median value (50th percentile) of the Bayestar photometric metallicity PDF & dex & float \\
\texttt{FeH\_Pristine\_bay\_84th} & Upper bound (84th percentile) of the Bayestar photometric metallicity PDF & dex & float \\
\texttt{mcfrac\_Pristine\_bay} & Fraction of MC samples inside model colour-colour space using Bayestar & & float \\
\texttt{flag\_good\_bay} & 1 if all quality cuts described in Section~\ref{sec:quality} are met for the Bayestar photometric metallicity & & int \\
\hline
\texttt{G} & \Gaia DR3 $G$ magnitude & mag & float \\
\texttt{BP} & \Gaia DR3 $G_{\rm BP}$ magnitude & mag & float \\
\texttt{RP} & \Gaia DR3 $G_{\rm RP}$ magnitude & mag & float \\
\texttt{d\_G} & uncertainty on the \Gaia $G$ magnitude & mag & float \\
\texttt{d\_BP} & uncertainty on the \Gaia $G_{\rm BP}$ magnitude & mag & float \\
\texttt{d\_RP} & uncertainty on the \Gaia $G_{\rm RP}$ magnitude & mag & float \\
\texttt{parallax} & \Gaia DR3 parallax & mas & float \\
\texttt{parallax\_error} & uncertainty on the \Gaia DR3 parallax & mas & float \\
\texttt{RUWE} & \Gaia DR3 Renormalised Unit Weight Error & & float \\
\texttt{Cstar} & \Gaia DR3 corrected flux excess, $C^*$, as defined in equation (6) of \citet{riello21} & & float \\
\texttt{Cstar\_1sigma} & Normalized standard deviation of $C^*$, as defined in equation (18) of \citet{riello21} & & float \\
\texttt{Pvar} & Probability for the source to be variable, derived from \Gaia DR3 $G$-band photometry & & float \\
\texttt{gvar} & \Gaia $G$-band fractional flux uncertainty times the square root of the number of $G$ observations & & float \\
\hline
\texttt{sel\_4MOST} & 1 if included in the 4MOST 4MIDABLE/\texttt{VMP\_PIGS} sub-survey target catalogue & & int \\
\hline
\end{tabular}
\end{table*}

\section{Additional figures}

\begin{figure*}
\centering
\includegraphics[width=1.0\hsize,trim={0.0cm 0.0cm 0.0cm 0.0cm}]{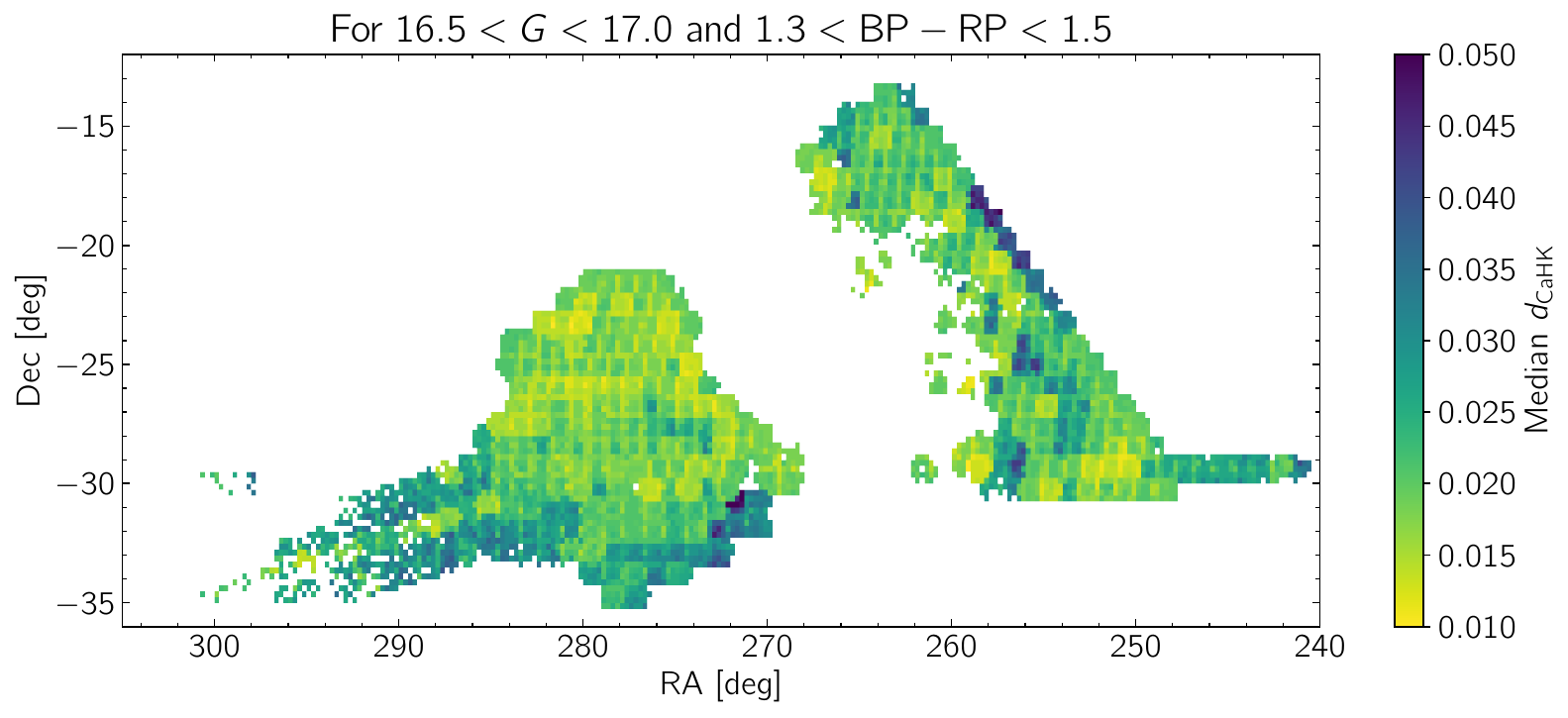}
\caption{Variation of survey depth across the PIGS footprint. Each 0.25 x 0.25 deg$^2$ pixel shows the median \CaHK uncertainty in that pixel for the magnitude and colour range indicated in the figure title. Pixels are included if there are at least 5 stars. Lighter narrow vertical stripes are due to overlaps between MegaCam fields, where combined magnitudes have lower uncertainties.
}
\label{fig:footdepth} 
\end{figure*}

\begin{figure*}
\centering
\includegraphics[width=0.7\hsize,trim={0.0cm 0.0cm 0.0cm 0.0cm}]{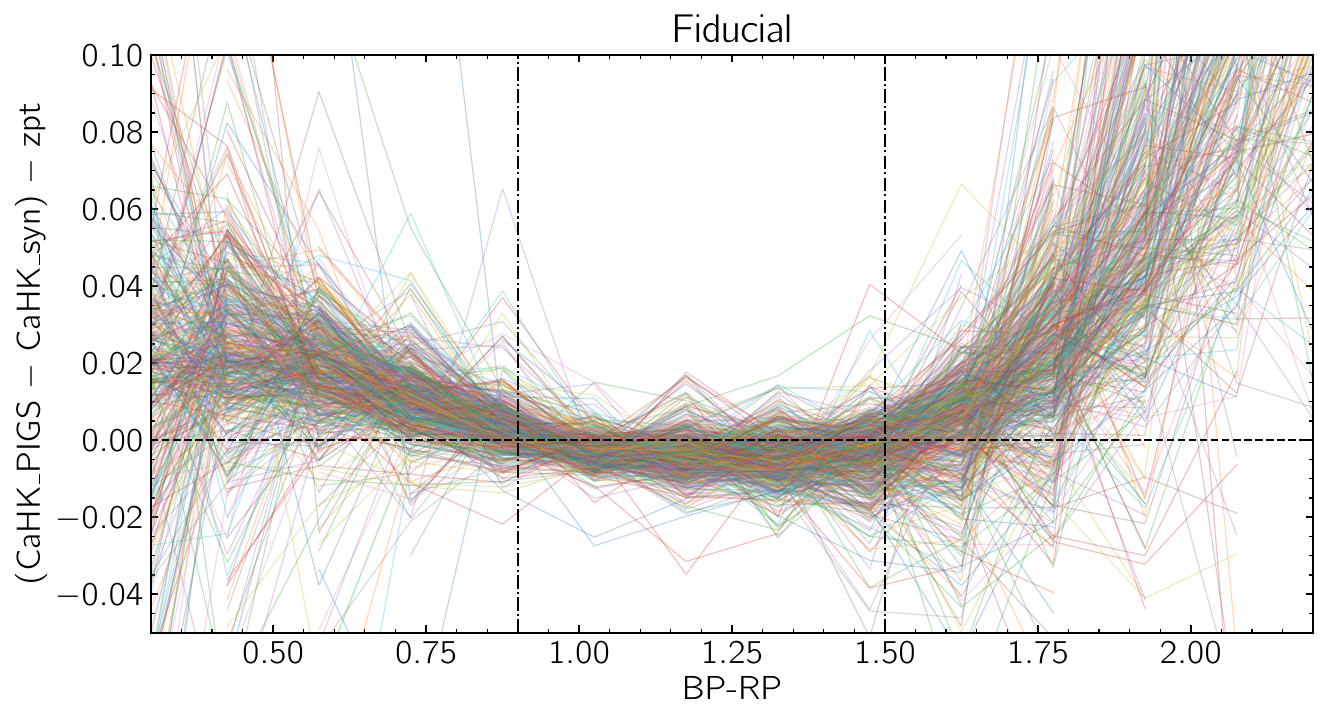}
\includegraphics[width=0.7\hsize,trim={0.0cm 0.0cm 0.0cm 0.0cm}]{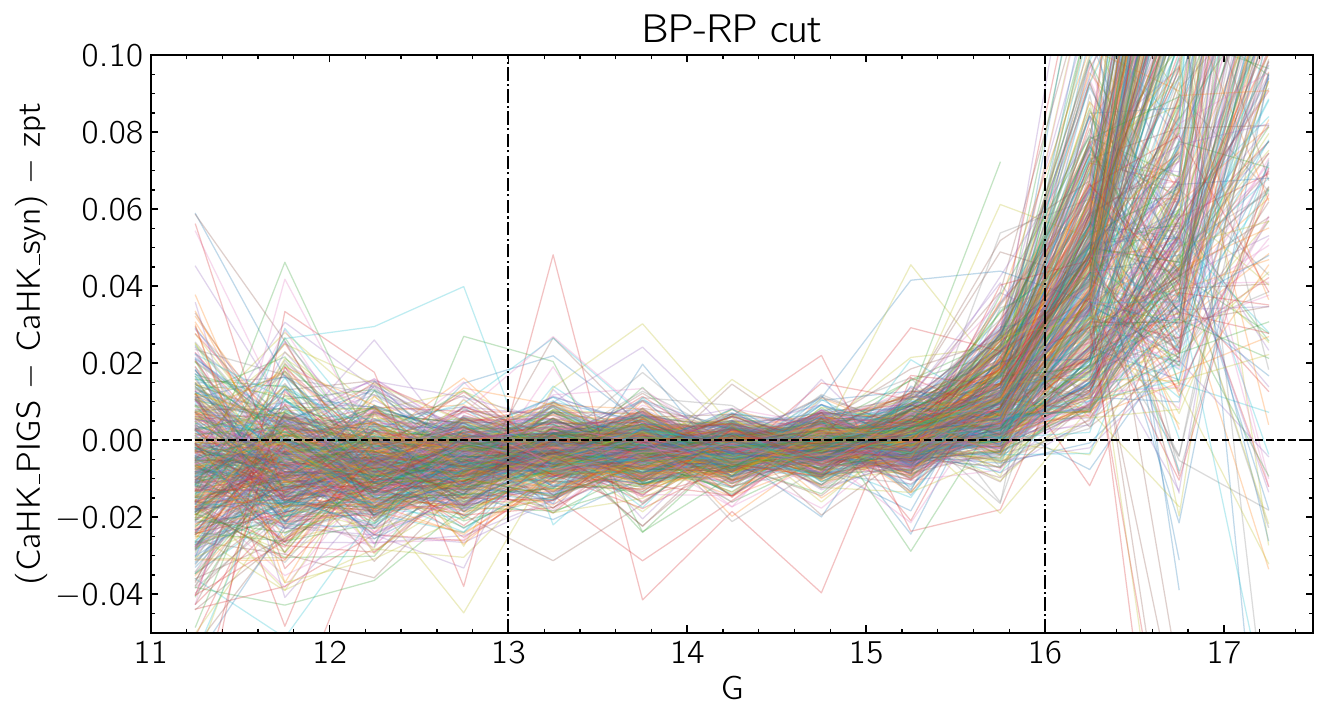}
\includegraphics[width=0.5\hsize,trim={0.0cm 0.0cm 0.0cm 0.0cm}]{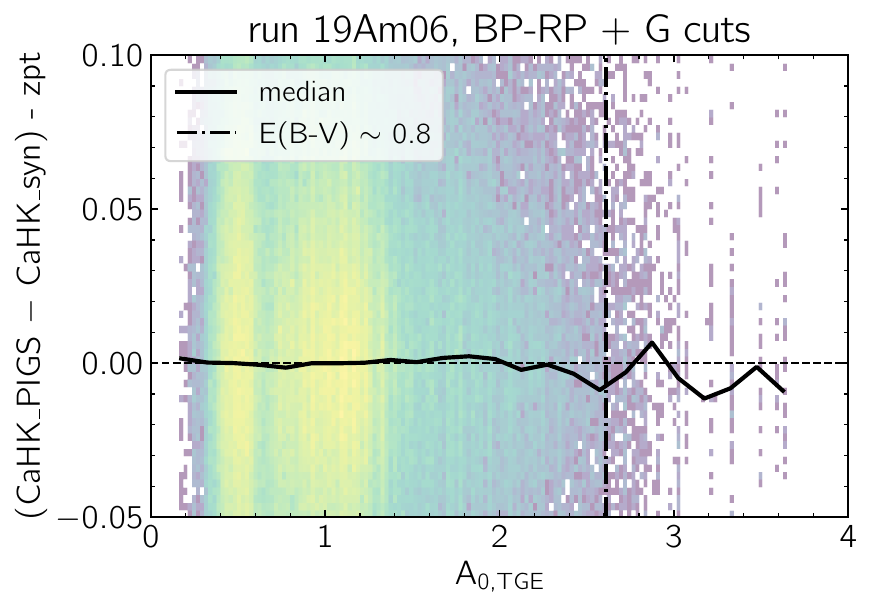}
\includegraphics[width=0.4\hsize,trim={0.0cm 0.0cm 0.0cm 0.0cm}]{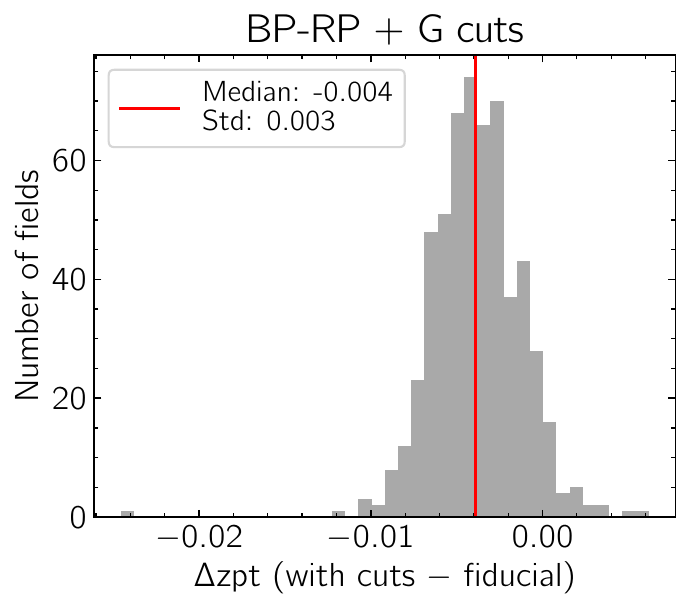}
\caption{Tests on the \CaHK calibration zero-point offsets following \citet{jplus}, who performed their photometric calibration using \Gaia XP synthetic photometry as functions of magnitude and colour. Top: for each MegaCam field, the difference between PIGS and synthetic \CaHK corrected by the derived zero-point as function of BP$-$RP, for the fiducial case as described in the main text. Middle: the same, but as function of $G$ and after cutting to the BP-RP regime shown with vertical lines in the first panel. Bottom left: after cutting to the BP$-$RP and $G$ regimes shown with vertical lines in the first and second panel, the \CaHK difference as function of A$_\mathrm{0,TGE}$ (scaled by 0.76) for one of the MegaCam runs. There is no remaining trend. Bottom right: difference in derived zero-points between the fiducial case and after cutting to the BP$-$RP and $G$ regimes shown with vertical lines in the first and second panel. The median and standard deviation are indicated. They are small with respect to the uncertainty floor derived in Section~\ref{sec:calib} and the variation in the colour-colour space (Figure~\ref{fig:ccd}), so they will not cause any significant deviations in derived metallicity.
}
\label{fig:cal2} 
\end{figure*}

\begin{figure*}
\centering
\includegraphics[width=1.0\hsize,trim={0.0cm 0.0cm 0.0cm 0.0cm}]{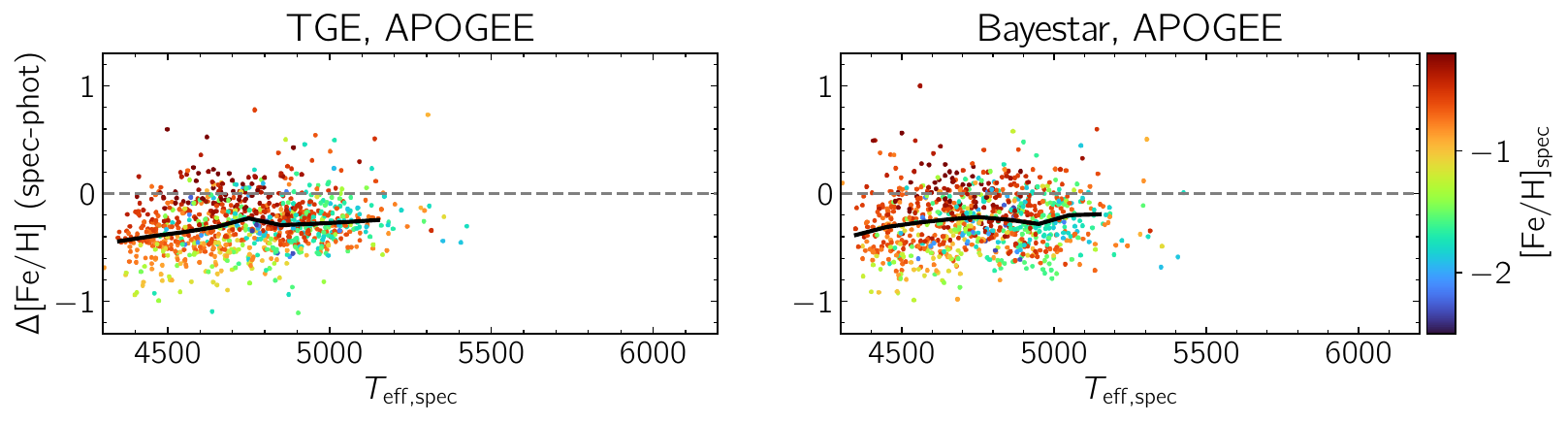}
\includegraphics[width=1.0\hsize,trim={0.0cm 0.0cm 0.0cm 0.0cm}]{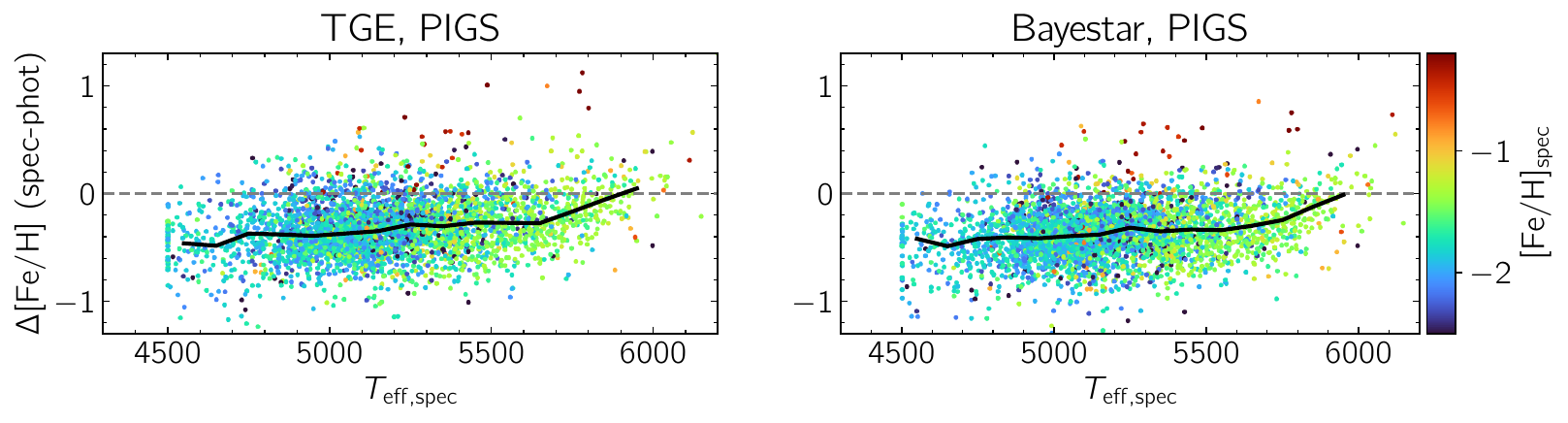}
\caption{Difference between photometric and spectroscopic metallicity (panel order as in Figure~\ref{fig:spec}) as function of spectroscopic \teff. Only stars with TGE E(B$-$V)$_\mathrm{TGE} <0.4$ and [Fe/H]$_\mathrm{phot,TGE} <-0.1$ have been included. A median trend line in bins of 100~K is shown in black. There is a slight trend with \teff, particularly for the lowest and highest \teff bins. In PIGS, stars at the highest temperatures are horizontal branch stars. In the APOGEE panels, there are a number of cool stars with $\feh \sim -1.7$ that are systematic outliers, which turn out to be dominated by stars in the M19 globular cluster -- specifically those closest to the cluster centre where crowding is the worst and the photometric metallicities may be expected to suffer. 
}
\label{fig:spec_teff} 
\end{figure*}

\begin{figure*}
\centering
\includegraphics[width=1.0\hsize,trim={0.0cm 0.0cm 0.0cm 0.0cm}]{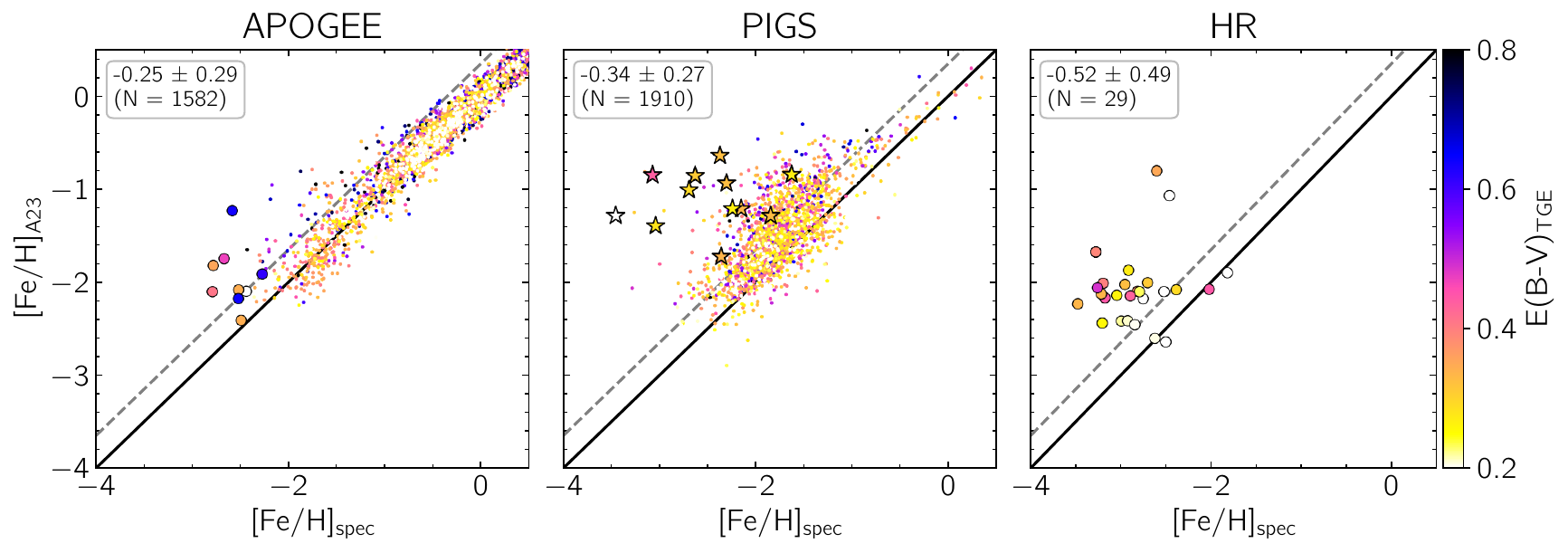}
\caption{Comparison of \citet{andrae23} \Gaia XP XGBoost vetted RGB metallicities with literature spectroscopic metallicities in the inner Galaxy, data and symbols as in Figure~\ref{fig:spec}. Note that not all the stars from Figure~\ref{fig:spec} are present in the \citetalias{andrae23} vetted RGB catalogue, due to their cuts on the magnitudes ($G<16$), parallaxes, XGBoost \teff \& \logg and the WISE \& \Gaia photometry. \textbf{Left:} APOGEE. The excellent agreement with APOGEE is largely thanks to those stars being included in the \citetalias{andrae23} training sample. \textbf{Middle:} PIGS. \citet{arentsen24} showed that those \citetalias{andrae23} metallicities that agree least well with the PIGS spectroscopy are typically faint in \Gaia BP, so XP results may improve with \Gaia DR4. \textbf{Right:} high-resolution spectroscopic literature. The \citetalias{andrae23} metallicities for extremely metal-poor stars appear to be significantly overestimated.
Numbers give the 3$\sigma$-clipped median and standard deviation of [Fe/H]$_\mathrm{spec} - $[Fe/H]$_\mathrm{phot}$, excluding [Fe/H]$_\mathrm{phot} > -0.1$.
}
\label{fig:a23_lit} 
\end{figure*}

\begin{figure*}
\centering
\includegraphics[width=1.0\hsize,trim={0.0cm 0.0cm 0.0cm 0.0cm}]{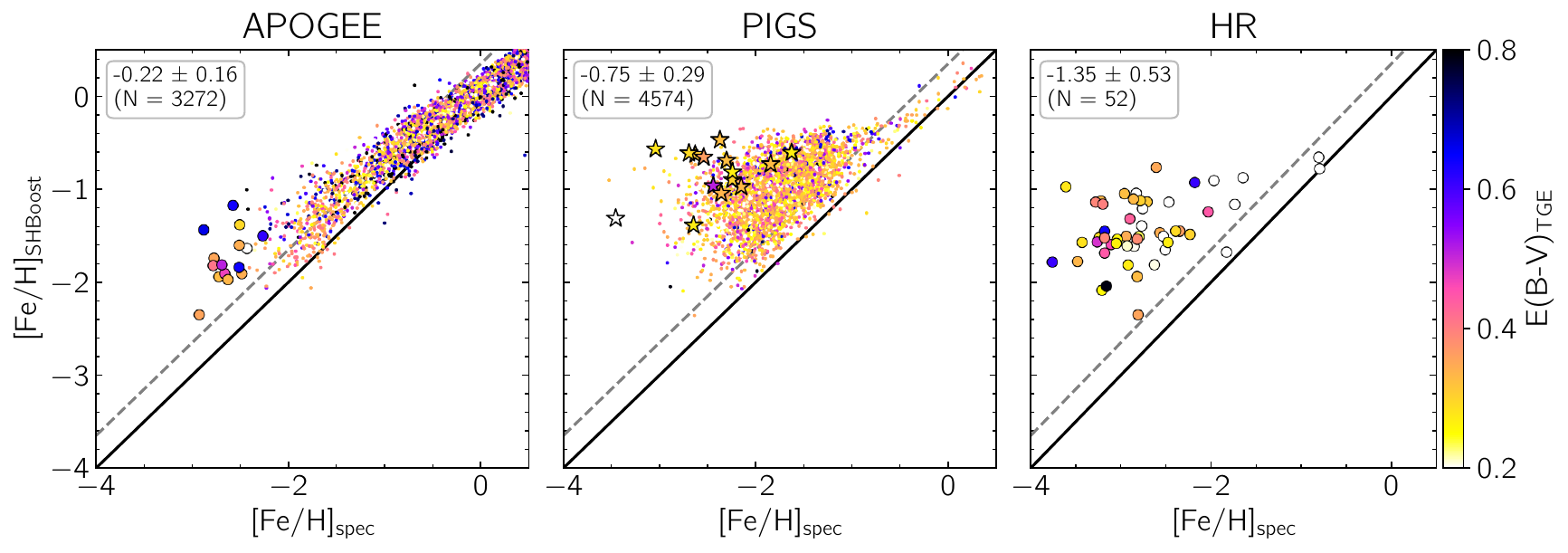}
\caption{Similar to Figure~\ref{fig:a23_lit}, but for the \citet{khalatyan24} \texttt{SHBoost} sample. Stars with quality flag $\texttt{xgb\_met\_outputflag} \neq 0$ have not been removed as that would exclude most VMP stars.
The stars in the PIGS sample are generally fainter than the APOGEE stars, which may contribute to the larger discrepancy with the PIGS metallicities. Additionally, APOGEE data was included in the training sample for \texttt{SHBoost}.
Numbers give the 3$\sigma$-clipped median and standard deviation of [Fe/H]$_\mathrm{spec} - $[Fe/H]$_\mathrm{phot}$, excluding [Fe/H]$_\mathrm{phot} > -0.1$.
}
\label{fig:sh_lit} 
\end{figure*}

\begin{figure*}
\centering
\includegraphics[width=0.7\hsize,trim={0.0cm 0.0cm 0.0cm 0.0cm}]{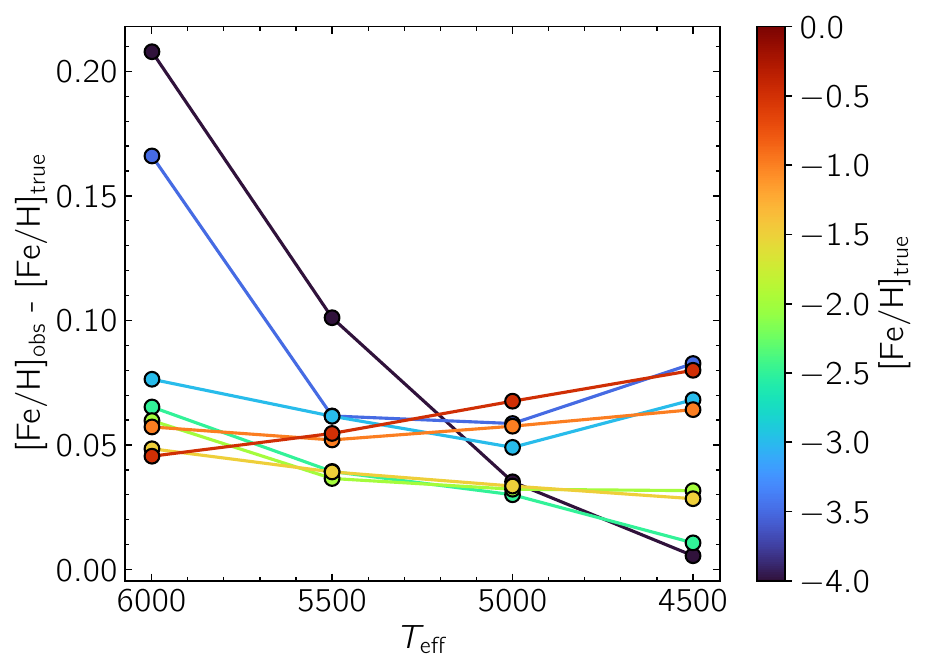}
\caption{The effect of interstellar Ca H\&K lines with a combined EW of 0.4~\AA on derived photometric metallicities ([Fe/H]$_\mathrm{obs}$) for a grid of synthetic spectra of giant stars with varying [Fe/H]$_\mathrm{true}$ and \teff and a fixed $\logg = 1.5$. The (BP$-$RP) was estimated by inverting the \Gaia infrared flux method of \citet{casagrande21} given the grid's \teff, [Fe/H]$_\mathrm{true}$ and \logg, and the y-axis of the \Pristine colour-colour space was estimated by inverting the photometric metallicity model given the above (BP$-$RP) and [Fe/H]$_\mathrm{true}$. The trends with \teff and \feh are a combination of the properties of the \Pristine colour-colour space and the effect of taking away flux from synthetic spectra with many or few absorption features already. 
The effect is strongest for the two most metal-poor and hottest stars. 
}
\label{fig:ismgrid} 
\end{figure*}

\bsp	%
\label{lastpage}
\end{document}